\documentclass[fleqn,usenatbib]{mnras}

\usepackage{amsmath,amssymb}
\usepackage{graphicx}
\usepackage{bm}
\hypersetup{colorlinks=true,linkcolor=blue,citecolor=blue,urlcolor=blue}

\newcommand{\figref}[2][]{\hyperref[#2]{Fig.~\ref*{#2}#1}}
\newcommand{\Figref}[2][]{\hyperref[#2]{Figure~\ref*{#2}#1}}
\newcommand{\tabref}[1]{\hyperref[#1]{Table~\ref*{#1}}}
\newcommand{\secref}[1]{\hyperref[#1]{Section~\ref*{#1}}}
\newcommand{\Secref}[1]{\hyperref[#1]{Section~\ref*{#1}}}
\newcommand{\appref}[1]{\hyperref[#1]{Appendix~\ref*{#1}}}
\newcommand{\Eqref}[1]{\hyperref[#1]{equation~(\ref*{#1})}}
\newcommand{\EQref}[1]{\hyperref[#1]{Equation~(\ref*{#1})}}
\newcommand{\Eqsref}[2]{\hyperref[#1]{equations~(\ref*{#1})}--\hyperref[#2]{(\ref*{#2})}}

\title[Relativistic precession around rotating neutron stars I]{Relativistic
precession around rotating neutron stars. I. Slow rotation: Hartle--Thorne
spacetimes, particle orbits and the QPOs of atoll and Z sources}

\author[A. C. Khunt]{A. C. Khunt$^{1}$\thanks{E-mail: ack.gravity@gmail.com}
\\
$^{1}$Department of Physics, Sardar Patel University, Vallabh Vidyanagar 388120, India.}

\begin{document}
\label{firstpage}
\pagerange{\pageref{firstpage}--\pageref{lastpage}}
\maketitle

\begin{abstract}
We study the relativistic precession model (RPM) of quasi-periodic
oscillations (QPOs) in the spacetimes of slowly rotating neutron stars
computed from the equation of state (EOS).  For four EOSs (WFF1, G3,
FSUGarnet and IOPB-I) we solve the Hartle--Thorne equations to second order
in the angular velocity and compute the orbital and epicyclic frequencies of
the exterior metric without expanding in the spin or the quadrupole.  At
$1.9\,\mathrm{M}_\odot$ the spin-induced quadrupole is 2--4 times that of a
black hole with the same spin; it lowers the nodal precession frequency by
up to 25 per cent near the innermost stable circular orbit and limits the
nodal frequency that any spin can produce.  An eccentric, tilted geodesic
around a $1.92\,\mathrm{M}_\odot$ star spinning at 363~Hz precesses at the
epicyclic rates, and the power spectrum of a hot spot on it peaks at the
orbital, periastron and nodal precession frequencies.  Applied to the atoll
sources 4U 1636$-$53 and 4U 1728$-$34, the RPM gives masses of
2.02--$2.15\,\mathrm{M}_\odot$ and 1.88--$1.92\,\mathrm{M}_\odot$.  For
4U 1728$-$34 the low-frequency QPO favours stars with large moments of
inertia, $I\simeq2.5\times10^{45}$~g\,cm$^{2}$.  Without an EOS the QPOs give
$M=1.920^{+0.014}_{-0.012}\,\mathrm{M}_\odot$, $j=0.162^{+0.023}_{-0.011}$ and
$\bar Q=2.2^{+1.1}_{-0.8}$; a Kerr spacetime with the same nodal frequency
would underestimate the moment of inertia by 22 per cent.  For the Z sources
GX 5$-$1 and GX 340+0 the horizontal-branch oscillations cannot be twice the
nodal precession frequency for any of the four EOSs: the fits are driven to
spins of 730--745~Hz and to the maximum mass.
\end{abstract}


\section{Introduction}
\label{sec:intro}

Low-mass X-ray binaries (LMXBs) that contain a weakly magnetized neutron
star are divided into atoll and Z sources according to the tracks they
trace in X-ray colour--colour diagrams and the variability that goes with
each part of the track~\citep{HasingerVanderKlis1989}.  Their power spectra
show quasi-periodic oscillations (QPOs) over a wide range of frequencies.
In 1996 the Rossi X-ray Timing Explorer (RXTE) found QPOs near 1~kHz in the
Z source Sco X-1~\citep{vanderKlis1996} and in the atoll sources
4U 1728$-$34~\citep{Strohmayer1996} and 4U 1608$-$52~\citep{Berger1996}.  These
kHz QPOs usually appear in pairs, the lower and upper kHz QPO, with
frequencies between a few hundred hertz and about 1.3~kHz, and both move
together as the source moves along its track~\citep{vanderKlis2000,vanderKlis2006}.
During thermonuclear bursts several of the same sources show nearly coherent
burst oscillations, which give the spin frequency of the star, for example
363~Hz in 4U 1728$-$34~\citep{Strohmayer1996} and 581~Hz in
4U 1636$-$53~\citep{StrohmayerMarkwardt2002}.  At lower frequencies, Z sources
show the horizontal-branch oscillation (HBO) at 15--60~Hz, and atoll sources
show QPOs and noise components with similar properties~\citep{vanderKlis2006}.  The frequencies of
the low-frequency features are correlated with those of the kHz QPOs over
a wide range, and similar correlations are found in black-hole
binaries~\citep{PsaltisBelloniVanderKlis1999}.  Since RXTE, kHz QPOs have also
been observed with the Large Area X-ray Proportional Counter (LAXPC) on
AstroSat~\citep{Chauhan2017,Anand2026}.  For 4U 1728$-$34, \citet{Anand2024} detected 13 simultaneous triplets of a low-frequency
QPO and twin kHz QPOs in a single observation, with the frequency of the
lower kHz QPO measured to about 1~Hz.

The kHz QPO frequencies are close to orbital frequencies a few gravitational
radii from a neutron star, and most models connect them with the motion of gas in
the inner accretion flow.  In beat-frequency models the upper kHz QPO is an
orbital frequency near the inner edge of the disc and the lower one its beat
with the stellar spin; such models were first proposed for the
HBO~\citep{AlparShaham1985,Lamb1985} and later developed for the kHz QPOs as
the sonic-point model~\citep{MillerLambPsaltis1998}.  The separation of the two
kHz QPOs, however, changes with their frequencies and is not simply related
to the spin~\citep{vanderKlis2006}.  Resonance models relate the two peaks to
a nonlinear coupling of epicyclic oscillations of the
flow~\citep{AbramowiczKluzniak2001}, and other models use oscillation modes of
thick tori~\citep{Matuszkova2024}.  The relativistic precession model (RPM) of
\citet{StellaVietri1998,StellaVietri1999} identifies the upper
kHz QPO with the orbital frequency, the lower kHz QPO with the periastron
precession frequency, and the low-frequency QPO with the nodal precession
frequency of a slightly eccentric and tilted orbit, all at the same radius.
To match the observed low frequencies, the last identification was later
made with twice the nodal frequency~\citep{StellaVietriMorsink1999}.  The
RPM uses only test-particle frequencies and has few free parameters.  It has been used to
estimate neutron-star masses~\citep{Torok2010,Torok2012,duBuisson2019,Anand2024}
and, for black holes, masses and spins~\citep{Motta2014,Motta2022}.  For the
low-frequency QPOs of black-hole binaries, Lense--Thirring precession of a hot
inner flow is a widely discussed interpretation~\citep{IngramDoneFragile2009,IngramMotta2019}.
The model also has known difficulties.  The RPM masses inferred for several
sources are high, about $2\,\mathrm{M}_\odot$ or more, and depend strongly on the
assumed spin~\citep{Torok2010,Torok2012}.  In a study of 13 sources with known spin,
\citet{vanDoesburgh2017} found that the
low-frequency QPO frequency rises more steeply with the upper kHz QPO
frequency than the RPM predicts, and that it shows no clear dependence on
the spin.

Of the three RPM frequencies, the nodal precession frequency is the most
sensitive to the spacetime of the star.  It contains the Lense--Thirring term, proportional to
the dimensionless spin $j=cJ/GM^{2}$, and a retrograde term caused by the
oblateness of the star, proportional to the quadrupole $q=c^{4}Q/G^{2}M^{3}$.
\citet{MorsinkStella1999} computed these frequencies for
numerical models of rapidly rotating neutron stars and showed that near the
star the two terms are comparable.  For a Kerr black hole $q=-j^{2}$.  For a
neutron star the reduced quadrupole $\bar Q=-q/j^{2}$ is several times
larger, and its value depends on the equation of state (EOS) and the
mass~\citep{LaarakkersPoisson1999,Pappas2012,YagiYunes2013}.  The orbital
frequencies around neutron stars have therefore been computed in
spacetimes with independent multipoles.  \citet{Pappas2012MNRAS} showed
how the multipole moments could be read from QPO frequencies in the RPM,
\citet{Pachon2012} compared the precession frequencies of
analytic exterior solutions with those of numerical stars, and \citet{Maselli2020} fitted kHz QPOs of three atoll sources, including
4U 1728$-$34, with an accurate analytic metric of rotating neutron stars.

The Hartle--Thorne (HT) metric~\citep{Hartle1967,HartleThorne1968} is the
simplest such spacetime.  It describes a rigidly rotating star to second
order in its angular velocity, its exterior depends only on $(M,J,Q)$, and
it agrees well with numerical models for the spin frequencies of most known
pulsars~\citep{Berti2005,Urbancova2019}.  \citet{Boshkayev2014,Boshkayev2015} derived the epicyclic frequencies
of the HT metric and used the kHz QPOs of atoll and Z sources to infer $M$,
$J$ and $Q$ as free parameters.  More recently they applied
the RPM in HT and other spacetimes to eight neutron-star
LMXBs~\citep{Boshkayev2024,Boshkayev2026}.  \citet{Urbancova2019} gave a systematic study of the orbital and epicyclic
frequencies of the HT metric for realistic EOSs, using the universal
relations between $I$, $Q$ and the stellar
parameters~\citep{YagiYunes2013}.  The same group has studied the oscillations
and precession of fluid tori in this metric~\citep{Matuszkova2024} and has
shown that, because of the competition between the Lense--Thirring and the
quadrupole terms, the nodal frequency at a given radius has a maximum as a
function of $j$, so that slowly and rapidly rotating stars can show the same
precession frequency~\citep{Torok2025}.  On the observational side, \citet{Anand2024} used the RPM in the Kerr metric to obtain
$M=1.92\pm0.01\,\mathrm{M}_\odot$ for 4U 1728$-$34 and a moment of inertia of about
$2\times10^{45}$~g\,cm$^{2}$, and estimated that a self-consistent neutron-star
metric would change these values by about 1~per~cent and 5~per~cent.

In most of these studies the spacetime is Kerr, or $j$ and $q$ are fitted
as independent parameters, or the low-frequency QPO is left out.
Studies with EOS-consistent stars, such as \citet{Urbancova2019}, have
mainly addressed the twin kHz QPOs.  Yet the EOS fixes $j$ and $q$ for a star of given mass and spin frequency,
and for the atoll sources the spin frequency is known from burst
oscillations.  Independent EOS
constraints now come from the masses of the heaviest
pulsars~\citep{Fonseca2021} and from NICER pulse-profile
modelling~\citep{Riley2021,Miller2021}.  In this paper we compute the spacetime
and the orbital frequencies together from the EOS.  For four EOSs we
solve the HT equations, including the spin-induced change of mass and the
deformed surface, and obtain $(j,q,R_{\rm eq})$ for any mass and spin.  We
compute the orbital and epicyclic frequencies from the full exterior metric
without expanding in $j$ or $q$.  We then integrate eccentric, tilted
geodesics around one of these stars and compute the power spectrum of a hot
spot moving on them, to see how the three RPM frequencies appear in the
flux.  Finally, we use the RPM to infer the masses and spins of the atoll
sources 4U 1636$-$53 and 4U 1728$-$34 and of the Z sources GX 5$-$1 and
GX 340+0.  We address three questions.  (i) How does the quadrupole of the star
change the precession frequencies compared with a black hole of the same
spin?  (ii) Do the precession frequencies of real orbits appear in the flux of
matter moving on them?  (iii) What masses and spins do the observed QPOs
imply, and can the HBOs of the Z sources be nodal precession?

This is the first paper of a series.  Here we treat the stars in the
slow-rotation approximation, which gives the multipoles and the frequencies
in closed form and makes their dependence on $j$ and $\bar Q$ transparent.
Rapidly rotating stars, computed with numerical spacetimes, will be treated
in a companion paper (Paper~II), which will also test the accuracy of the
approximation used here.

\Secref{sec:ht} summarises the HT stars and \secref{sec:freq} the
orbital frequencies and the RPM.  \Secref{sec:stars} presents the
stellar and orbital properties for the four EOSs, and
\secref{sec:orbits} the particle orbits and their power spectra.
\Secref{sec:obs} confronts the model with the observed QPOs, and
\secref{sec:discussion} discusses the results.  The appendices give the
HT equations and the derivation of the epicyclic frequencies.  We use
$G=c=1$ except where frequencies are quoted in Hz.

\section{Slowly rotating stars in the Hartle--Thorne approximation}
\label{sec:ht}

Consider a star of central energy density $\varepsilon_c$ that rotates rigidly
with angular velocity $\Omega=2\pi\nu_s$.  \citet{Hartle1967} expanded the metric to
second order in $\Omega$ about the nonrotating Tolman--Oppenheimer--Volkoff
(TOV) solution,
\begin{align}
ds^{2}=&-e^{\nu}\left[1+2\left(h_{0}+h_{2}P_{2}\right)\right]dt^{2}\nonumber\\
&+e^{\lambda}\left[1+\frac{2\left(m_{0}+m_{2}P_{2}\right)}{r-2m}\right]dr^{2}\nonumber\\
&+r^{2}\left[1+2k_{2}P_{2}\right]\left[d\theta^{2}+\sin^{2}\theta\left(d\phi-\omega\,dt\right)^{2}\right],
\label{eq:htmetric}
\end{align}
where $P_{2}=P_{2}(\cos\theta)$ is the Legendre polynomial, $e^{\lambda}=(1-2m/r)^{-1}$, and
$\nu(r)$, $m(r)$ are the TOV potential and mass function.  The function
$\omega(r)$ is of order $\Omega$, and $h_{0}$, $m_{0}$, $h_{2}$, $m_{2}$,
$k_{2}$ are of order $\Omega^{2}$.  We first solve the TOV equations and
then three perturbation problems in turn.  The equations are collected in
\appref{app:interior}; here we summarise what each one gives.

The first problem ($\ell=1$) describes frame dragging.  The angular velocity
of the fluid relative to the local inertial frames, $\varpi=\Omega-\omega$, obeys a second-order
equation that is regular at the centre.  Matching to the exterior solution
$\omega=2J/r^{3}$ gives the angular momentum $J$ and the moment of inertia
$I=J/\Omega$.

The second problem ($\ell=0$) describes the spherical deformation.  The
functions $m_{0}$ and $p_{0}^{*}$
give the change of the gravitational mass at fixed central density,
$\delta M=m_{0}(R)+J^{2}/R^{3}$, where $R$ is the TOV radius.  The mass that
enters the exterior metric and all orbital frequencies is the total mass
$M=M_{\rm TOV}+\delta M$.  All masses quoted in this paper are total masses
of rotating stars.

The third problem ($\ell=2$) describes the quadrupolar deformation.  The
functions $h_{2}$ and $v_{2}=h_{2}+k_{2}$ obey two coupled first-order
equations.  Matching them to the exterior solution fixes a constant $K$ and
the quadrupole moment
\begin{equation}
Q=-\left(\frac{J^{2}}{M}+\frac{8}{5}KM^{3}\right).
\label{eq:Q}
\end{equation}
We use the Geroch--Hansen sign convention, in which an oblate star has $Q<0$
and the Kerr metric has $Q=-J^{2}/M$.  The dimensionless spin and
quadrupole are $j=J/M^{2}$ and $q=Q/M^{3}$.  We also use the reduced
quadrupole
\begin{equation}
\bar Q\equiv-\frac{q}{j^{2}}=-\frac{QM}{J^{2}},
\end{equation}
which equals unity for a Kerr black hole and lies between about 1.5 and 12
for the neutron stars considered here (\figref{fig:stars}).  In the Hartle--Thorne (HT)
approximation $J\propto\Omega$ and $Q,\delta M\propto\Omega^{2}$ at fixed
central density, so to leading order $\bar Q$ depends on the EOS and the mass but
not on the spin.

The surface of the rotating star is
$r(\theta)=R+\xi_{0}+\xi_{2}P_{2}(\cos\theta)$, with $\xi_{0}$ and $\xi_{2}$ from
the $\ell=0$ and $\ell=2$ pressure perturbations (\appref{app:interior}).
The equatorial coordinate radius $R_{\rm eq}=R+\xi_{0}-\xi_{2}/2$ is the radius
at which an accretion disc would touch the star, in the same coordinate as
the exterior metric.  The circumferential equatorial radius is
$R_{\rm eq,circ}=R_{\rm eq}(1-k_{2}/2)$, which is the quantity measured by
pulse-profile modelling.

Outside the star the metric functions are known in closed
form~\citep{HartleThorne1968}.  With $\xi=r/M-1$ and the associated Legendre
functions of the second kind $Q_{2}^{1}$, $Q_{2}^{2}$ (\appref{app:exterior}),
\begin{align}
\omega&=\frac{2J}{r^{3}},\qquad h_{0}=\frac{J^{2}}{r^{3}(r-2M)},\qquad m_{0}=-\frac{J^{2}}{r^{3}},\nonumber\\
h_{2}&=\frac{J^{2}}{Mr^{3}}\left(1+\frac{M}{r}\right)+K\,Q_{2}^{2}(\xi),\nonumber\\
v_{2}&=-\frac{J^{2}}{r^{4}}+K\frac{2M}{\sqrt{r(r-2M)}}\,Q_{2}^{1}(\xi),
\label{eq:exterior}
\end{align}
with $m_{2}=(r-2M)(-h_{2}+6J^{2}/r^{4})$ and $k_{2}=v_{2}-h_{2}$.  The metric
depends on $(M,J,Q)$ only, so for given $(M,j,q)$ the exterior and all
orbital frequencies are fixed, whatever the interior.  The EOS enters through
the map $(M,\nu_s)\mapsto(j,q,R_{\rm eq})$.  Because the metric is correct to
$\mathcal{O}(\Omega^{2})$, setting $q=-j^{2}$ reproduces the Kerr metric to
$\mathcal{O}(j^{2})$, which is a useful check of the frequencies
(\appref{app:epicyclic}).

\section{Orbital frequencies and the relativistic precession model}
\label{sec:freq}

\subsection{Circular orbits and epicyclic frequencies}

Consider a test particle on a prograde circular orbit of radius $r$ in the
equatorial plane of a stationary, axisymmetric, reflection-symmetric metric
with components $g_{tt}$, $g_{t\phi}$, $g_{\phi\phi}$, $g_{rr}$, $g_{\theta\theta}$.  The
coordinate angular velocity $\Omega_{\phi}=d\phi/dt$ follows from the radial
geodesic equation,
\begin{equation}
\Omega_{\phi}=\frac{-g_{t\phi,r}+\sqrt{g_{t\phi,r}^{2}-g_{tt,r}\,g_{\phi\phi,r}}}{g_{\phi\phi,r}},
\label{eq:omphi}
\end{equation}
and the normalisation of the four-velocity gives
$u^{t}=[-(g_{tt}+2\Omega_{\phi}g_{t\phi}+\Omega_{\phi}^{2}g_{\phi\phi})]^{-1/2}$, the
energy $E=-(g_{tt}+g_{t\phi}\Omega_{\phi})u^{t}$ and the angular momentum
$L=(g_{t\phi}+g_{\phi\phi}\Omega_{\phi})u^{t}$ per unit mass.  For fixed $E$ and $L$
the normalisation $u\cdot u=-1$ reads
\begin{align}
g_{rr}\dot r^{2}+g_{\theta\theta}\dot\theta^{2}&=-1-\mathcal{H}(r,\theta),\nonumber\\
\mathcal{H}&=\frac{g_{\phi\phi}E^{2}+2g_{t\phi}EL+g_{tt}L^{2}}{g_{tt}g_{\phi\phi}-g_{t\phi}^{2}},
\label{eq:H}
\end{align}
where the dot denotes the derivative with respect to proper time.
Linearising about the circular orbit gives harmonic oscillations in $r$ and
$\theta$ with proper-time frequencies $\kappa_{r}^{2}=\mathcal{H}_{,rr}/2g_{rr}$ and
$\kappa_{\theta}^{2}=\mathcal{H}_{,\theta\theta}/2g_{\theta\theta}$ (\appref{app:epicyclic}).  The
frequencies seen by a distant observer are
\begin{equation}
\nu_{\phi}=\frac{\Omega_{\phi}}{2\pi},\qquad
\nu_{r}=\frac{\kappa_{r}}{2\pi u^{t}},\qquad
\nu_{\theta}=\frac{\kappa_{\theta}}{2\pi u^{t}}.
\label{eq:nus}
\end{equation}
The innermost stable circular orbit (ISCO) is the outermost zero of
$\kappa_{r}^{2}$.  The disc ends at
$r_{\rm in}=\max(R_{\rm eq},r_{\rm ISCO})$, that is, at the ISCO or at the
stellar surface, whichever lies farther out.
We evaluate \Eqsref{eq:omphi}{eq:nus} for the full HT metric of
\Eqref{eq:exterior}, with fourth-order finite differences.  We do not
expand in $j$ or $q$, so the frequencies are those of the metric in which we
integrate the particle orbits (\secref{sec:orbits}).

\subsection{Periastron and nodal precession}

The differences
\begin{equation}
\nu_{\rm per}=\nu_{\phi}-\nu_{r},\qquad \nu_{\rm nod}=\nu_{\phi}-\nu_{\theta}
\end{equation}
are the periastron and nodal precession frequencies of a slightly eccentric,
slightly tilted orbit.  Their weak-field structure explains most of what
follows.  To leading order in each multipole (\appref{app:weak}),
\begin{align}
\frac{\nu_{r}^{2}}{\nu_{\phi}^{2}}&\simeq1-\frac{6}{x}+\frac{8j}{x^{3/2}}+\frac{3q}{x^{2}},\nonumber\\
\frac{\nu_{\theta}^{2}}{\nu_{\phi}^{2}}&\simeq1-\frac{4j}{x^{3/2}}-\frac{3q}{x^{2}},
\label{eq:weak}
\end{align}
with $x=r/M$.  For $q=-j^{2}$ these reduce to the familiar Kerr
expressions~\citep{Okazaki1987,StellaVietri1998}, and for $j=0$ they give the
Newtonian precession in the field of a mass quadrupole.  The nodal precession
is then
\begin{align}
\nu_{\rm nod}&\simeq\nu_{\rm LT}+\nu_{Q},\nonumber\\
\nu_{\rm LT}&=\frac{\omega}{2\pi}=\frac{GJ}{\pi c^{2}r^{3}},\qquad
\nu_{Q}=\frac{3}{2}\frac{q}{x^{2}}\,\nu_{\phi}.
\label{eq:nod}
\end{align}
The first term is the Lense--Thirring precession, equal to the frame-dragging
angular velocity of zero-angular-momentum observers.  The second is the
classical precession driven by the oblateness of the star.  It is retrograde
($q<0$) and grows as $\nu_s^{2}$, while $\nu_{\rm LT}$ grows as $\nu_s$.  Their
ratio is
\begin{equation}
\frac{\nu_{Q}}{\nu_{\rm LT}}=-\frac{3}{4}\,\frac{\bar Q\,j}{\sqrt{x}},
\label{eq:ratio}
\end{equation}
which is about $-0.1$ to $-0.3$ near the ISCO for the $1.9\,\mathrm{M}_\odot$ stars of
\secref{sec:stars} at $\nu_s=363$--581~Hz.  The quadrupole is therefore not a
small correction to the nodal frequency, and replacing the star by a Kerr
metric with the same $j$ ($\bar Q=1$) overestimates $\nu_{\rm nod}$.  The
periastron precession is dominated by the $6/x$ term and is much less
sensitive to $j$ and $q$.

\subsection{The relativistic precession model}

In the RPM~\citep{StellaVietri1998,StellaVietri1999,StellaVietriMorsink1999} the
three QPOs of a source come from the same radius,
\begin{equation}
\nu_{U}=\nu_{\phi},\qquad \nu_{L}=\nu_{\rm per},\qquad \nu_{\rm LF}=2\nu_{\rm nod},
\label{eq:rpm}
\end{equation}
where $\nu_{U}$ and $\nu_{L}$ are the upper and lower kHz QPOs.  The
low-frequency QPO $\nu_{\rm LF}$ is the 20--50~Hz QPO of atoll sources or the
horizontal-branch oscillation (HBO) of Z sources.  The factor 2 in the last
relation was introduced by \citet{StellaVietriMorsink1999}
because the nodal frequency alone falls short of the observed low
frequencies.  It can be justified if the modulation has two maxima per
precession cycle.  Sets of QPOs observed at different times come from
different radii; once $(M,j,q)$ are fixed, this radius is the only freedom
of the model.  \Secref{sec:orbits} shows how the three
frequencies appear in the flux of matter on a real orbit.

The pair $(\nu_{U},\nu_{L})$ mainly constrains the mass, because
$\nu_{\rm per}/\nu_{\phi}$ depends mainly on $x$, and $\nu_{\phi}$ then fixes $M$.  The
low-frequency QPO adds information on the combination
$j(1-\tfrac{3}{4}\bar Q j/\sqrt{x})$ [\Eqsref{eq:nod}{eq:ratio}].  With a known
spin frequency, $j=2\pi\nu_s I/M^{2}$, so the low-frequency QPO measures a
combination of the moment of inertia and the quadrupole.  This is where an
EOS-consistent spacetime matters.

\section{Equations of state and stellar properties}
\label{sec:stars}

\subsection{Equations of state}

We use four EOSs that give radii of 10.5--13.3~km at $1.4\,\mathrm{M}_\odot$ and
maximum masses of 2.0--$2.15\,\mathrm{M}_\odot$ (\tabref{tab:eos}).  WFF1~\citep{WFF1988} is a microscopic
variational EOS (AV14 two-nucleon plus Urbana VII three-nucleon
interactions), matched to the Negele--Vautherin crust; we use the table
distributed with the RNS code~\citep{Stergioulas1995}.  It is soft at
intermediate densities, with $R_{1.4}=10.5$~km, but still reaches
$M_{\max}=2.15\,\mathrm{M}_\odot$.  Its radius lies below the range inferred from NICER
data~\citep{Miller2021}, and we keep it as a limiting compact case.  G3~\citep{Kumar2017G3}, FSUGarnet~\citep{ChenPiekarewicz2015}
and IOPB-I~\citep{Kumar2018IOPB} are relativistic mean-field (RMF) EOSs
calibrated to finite nuclei.  We use the unified versions of \citet{Parmar2022}, in which the inner crust is computed with the same
functional as the core and the outer crust from measured and
Hartree--Fock--Bogoliubov nuclear masses; the tables are available at
\url{https://github.com/hcdas/Unified_eos}.  They give $R_{1.4}=12.6$, 13.1
and 13.3~km and $M_{\max}=2.00$, 2.07 and $2.15\,\mathrm{M}_\odot$.  All four reach about $2\,\mathrm{M}_\odot$, as required by the
heaviest radio pulsars~\citep{Fonseca2021}, but none reaches $2.2\,\mathrm{M}_\odot$
without rotation.  Only the relation $\varepsilon(P)$ enters the structure
equations.

\begin{table*}
\caption{\label{tab:eos}Stellar properties for the four EOSs.  $M_{\max}$ is the maximum mass of nonrotating stars, $R_{1.4}$ and $R_{1.9}$ are nonrotating radii, and $\bar I_{1.4}=I/M^{3}$ is given at $1.4\,\mathrm{M}_\odot$.  $I$, $j$, $\bar Q=-q/j^{2}$ and $R_{\rm eq}/M$ are given at $M=1.9\,\mathrm{M}_\odot$ and $\nu_s=363$~Hz.}
\begin{tabular}{lcccccccc}
\hline
EOS & $M_{\max}\,[\mathrm{M}_\odot]$ & $R_{1.4}$ [km] & $R_{1.9}$ [km] & $\bar I_{1.4}$ & $I_{1.9}\,[10^{45}\,{\rm g\,cm^2}]$ & $j_{1.9}$ & $\bar Q_{1.9}$ & $R_{\rm eq}/M$ \\
\hline
WFF1 & 2.145 & 10.46 & 10.28 & 9.67 & 1.81 & 0.130 & 2.09 & 3.68 \\
G3 & 1.998 & 12.56 & 11.87 & 12.56 & 2.12 & 0.152 & 2.77 & 4.27 \\
FSUGarnet & 2.067 & 13.08 & 12.75 & 13.56 & 2.40 & 0.172 & 3.37 & 4.60 \\
IOPB-I & 2.150 & 13.31 & 13.08 & 13.88 & 2.50 & 0.179 & 3.58 & 4.72 \\
\hline
\end{tabular}
\end{table*}

\subsection{Spin, quadrupole and the inner edge of the disc}

\begin{figure}
\includegraphics[width=\columnwidth]{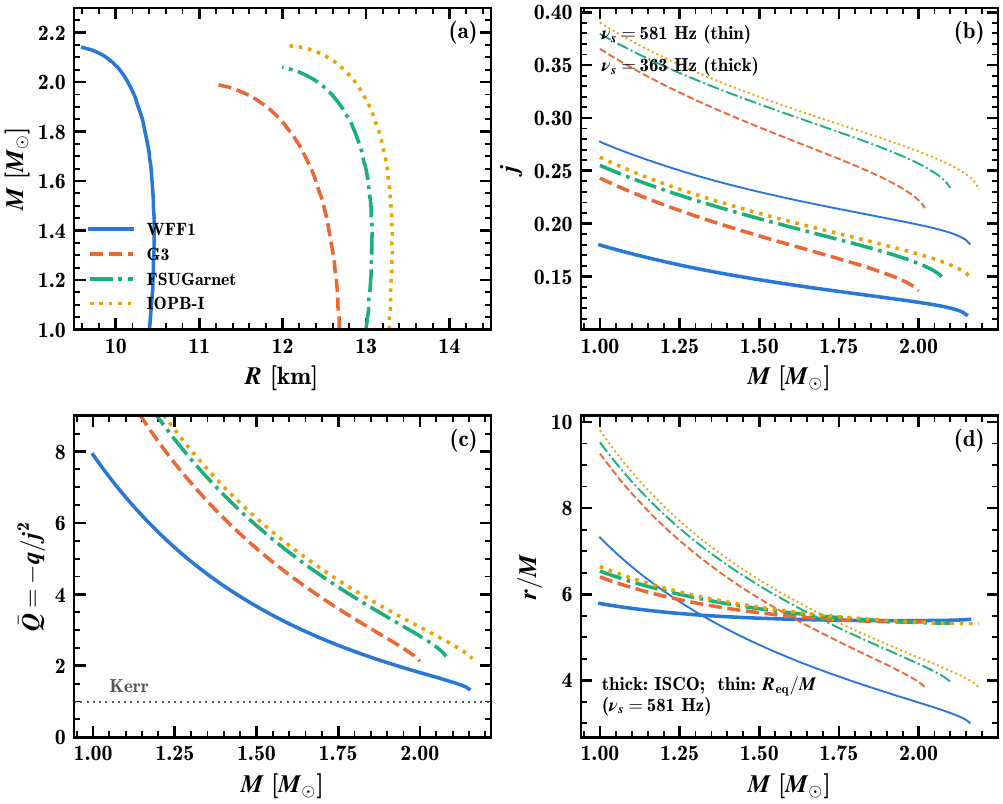}
\caption{\label{fig:stars}Stellar properties for the four EOSs.
(a) Mass--radius relations of nonrotating stars.  (b) Dimensionless spin
$j=cJ/GM^{2}$ at $\nu_s=363$~Hz (thick) and 581~Hz (thin).  (c) Reduced
quadrupole $\bar Q=-q/j^{2}$ at 363~Hz; the dotted line is the Kerr value.
(d) ISCO radius (thick) and equatorial radius (thin) in units of $M$ at
581~Hz.  The disc ends at the larger of the two.}
\end{figure}

\Figref{fig:stars} shows the properties that enter the orbital
frequencies.  At fixed mass and spin, $j$ grows with the stellar radius,
because $j=2\pi\nu_sI/M^{2}$ and $I\sim MR^{2}$.  At $1.9\,\mathrm{M}_\odot$ and 363~Hz,
$j$ ranges from 0.130 (WFF1) to 0.179 (IOPB-I).  The reduced quadrupole
$\bar Q$ falls with compactness, from about 12 at $1\,\mathrm{M}_\odot$ for the RMF
EOSs to 2--3.6 at $1.9\,\mathrm{M}_\odot$, and it is always larger than the Kerr value
$\bar Q=1$.  At $1.9\,\mathrm{M}_\odot$ the quadrupole $q=-\bar Qj^{2}$ is thus 2--4 times
larger in magnitude than for a black hole with the same $j$.  When $\bar I=I/M_{0}^{3}$ and $\bar Q=-QM_{0}/J^{2}$ are formed
with the nonrotating mass $M_{0}$, our stars follow the EOS-insensitive
$\bar I$--$\bar Q$ relation of \citet{YagiYunes2013} to 0.5--1.3~per~cent, which is
a check of the $\ell=1$ and $\ell=2$ solutions.  In the exterior metric, and
everywhere else in this paper, $j$ and $q$ are normalised by the total mass
$M$; the two normalisations differ at order $\delta M/M$.

The rotation increases the mass at fixed central density.  For a star of
total mass $1.4\,\mathrm{M}_\odot$, $\delta M/M=1$--2~per~cent at 363~Hz and 3--6~per~cent at 581~Hz (WFF1
to IOPB-I), and about half of that at $1.9\,\mathrm{M}_\odot$.  The maximum mass grows by
only 0.3--0.8~per~cent at 363~Hz (to 2.15, 2.01, 2.08 and $2.17\,\mathrm{M}_\odot$), because
compact stars are less deformed.  At a given total mass a rotating star
therefore has a lower central density than a nonrotating one, and a larger
radius.  This matters because the fits place several sources close to the maximum
mass.

The inner edge of the disc is the ISCO when the star is compact enough.  In
\figref[(d)]{fig:stars} the ISCO lies at $x\simeq5.3$--5.8 and depends weakly on
the EOS through $j$ and $q$.  The equatorial radius $R_{\rm eq}/M$ falls steeply
with mass.  At 581~Hz the star extends beyond the ISCO below
$M=1.32\,\mathrm{M}_\odot$ (WFF1), 1.59 (G3), 1.67 (FSUGarnet) and $1.70\,\mathrm{M}_\odot$ (IOPB-I).
In those cases the disc reaches the stellar surface and the highest
orbital frequency is the Keplerian frequency at $R_{\rm eq}$.  For the
masses preferred by the fits below, the inner edge is always the ISCO.

\subsection{Orbital frequencies}

\begin{figure*}
\includegraphics[width=0.9\textwidth]{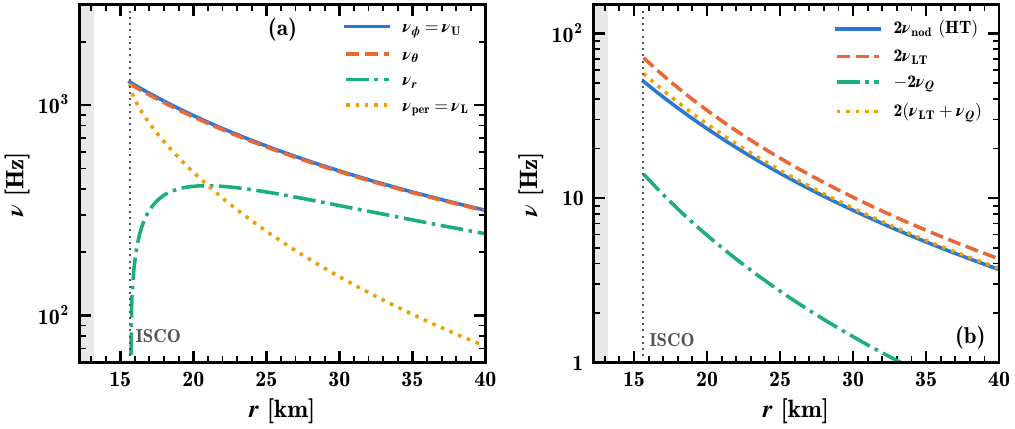}
\caption{\label{fig:freqs}Orbital frequencies around an IOPB-I star of
$1.922\,\mathrm{M}_\odot$ spinning at 363~Hz.  (a) Azimuthal, vertical and radial
epicyclic frequencies and the periastron precession frequency (the RPM
$\nu_{U}$ and $\nu_{L}$).  (b) Twice the nodal precession frequency (the RPM
$\nu_{\rm LF}$), the Lense--Thirring term $2\nu_{\rm LT}=\omega/\pi$, the
(retrograde) quadrupole term $-2\nu_{Q}$ and the weak-field sum,
\Eqref{eq:nod}.  The dotted vertical line is the ISCO and the shaded
band is the interior of the star.}
\end{figure*}

\Figref{fig:freqs} shows the frequencies of \secref{sec:freq} for an
IOPB-I star of $1.922\,\mathrm{M}_\odot$ at 363~Hz ($j=0.178$, $\bar Q=3.47$).  This is
the preferred model for 4U 1728$-$34 (\secref{sec:obs}).  The ISCO is at
15.65~km ($x=5.51$), 2.4~km outside the stellar equator, where $\nu_{\phi}=1285$~Hz.  The
vertical frequency $\nu_{\theta}$ is within 2~per~cent of $\nu_{\phi}$.  The radial
frequency $\nu_{r}$ has a maximum of 414~Hz at $x\approx7.4$ and vanishes at
the ISCO, so $\nu_{\rm per}=\nu_{L}$ rises from 0 at large radii to $\nu_{\phi}$ at the
ISCO.  Panel (b) shows $2\nu_{\rm nod}$ together with the
Lense--Thirring term $2\nu_{\rm LT}$, the quadrupole term $2\nu_{Q}$ and their
weak-field sum, \Eqref{eq:nod}.  Near the ISCO the exact nodal frequency
is 28~per~cent below the Lense--Thirring term.  The weak-field sum recovers most of
this difference.  It overestimates the exact value by 12~per~cent at the ISCO and
by 3~per~cent at $x=10$; the rest is due to higher-order terms in $1/x$.

\Figref[(a)]{fig:nodal} compares the HT nodal frequency with that of a Kerr
metric with the same $M$ and $j$.  At $1.9\,\mathrm{M}_\odot$ and 363~Hz the HT value is
lower by 3--7~per~cent (WFF1), 7--14~per~cent (G3), 10--22~per~cent (FSUGarnet) and 11--25~per~cent
(IOPB-I) between $x=14$ and the ISCO.  The deficit grows
with the radius of the star, because both $j$ and $\bar Q$ grow.  Panel (b)
shows a less familiar consequence.  At a fixed upper kHz QPO frequency
($\nu_{U}=900$~Hz) the nodal frequency first grows linearly with the spin
frequency, as $\nu_{\rm LT}\propto\nu_s$, but the quadrupole term grows as
$\nu_s^{2}$.  For the stiffer RMF EOSs the two balance at 600--650~Hz, and
$2\nu_{\rm nod}$ reaches a maximum of 32--34~Hz.  For the more compact G3 and WFF1
stars it is still rising at 750~Hz, the highest spin shown, where it
reaches 38--40~Hz.  A fast spin alone therefore cannot produce a high nodal
frequency.  \citet{Torok2025} found the same behaviour at
fixed radius: the precession frequency has a maximum as a function of $j$.
The height of the maximum depends on the mass and, through $I$ and $\bar Q$,
on the EOS.  It decides the interpretation of the Z sources in
\secref{sec:obs}.

\begin{figure*}
\includegraphics[width=0.9\textwidth]{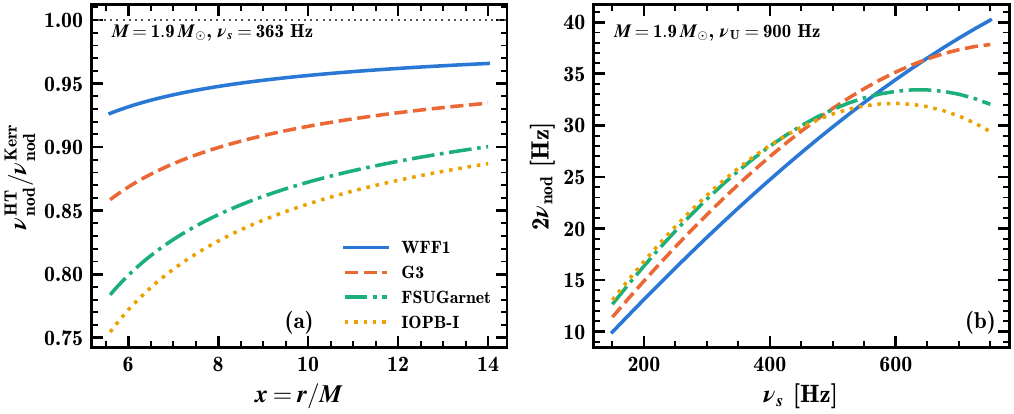}
\caption{\label{fig:nodal}(a) Ratio of the nodal precession frequency in
the HT metric to that in a Kerr metric with the same $M$ and $j$, for
$M=1.9\,\mathrm{M}_\odot$ and $\nu_s=363$~Hz.  (b) $2\nu_{\rm nod}$ at the radius where
$\nu_{\phi}=900$~Hz as a function of the spin frequency, for $M=1.9\,\mathrm{M}_\odot$.
The spin-induced quadrupole limits the nodal frequency that a star can
produce.}
\end{figure*}

\section{Particle orbits and QPO peaks}
\label{sec:orbits}

The epicyclic frequencies of \secref{sec:freq} describe small
oscillations about a circular orbit.  In this section we follow a full
geodesic of finite amplitude and ask how its frequencies would appear in
the X-ray flux.

\subsection{An eccentric, tilted orbit}

We take the IOPB-I star of \figref{fig:freqs} ($M=1.922\,\mathrm{M}_\odot$,
$\nu_s=363$~Hz, $j=0.178$, $\bar Q=3.47$, $R_{\rm eq}=13.21$~km, ISCO at
15.65~km), which is the preferred model for 4U 1728$-$34
(\secref{sec:obs}).  We integrate the geodesic equation
$\ddot x^{\mu}=-\Gamma^{\mu}_{\alpha\beta}\dot x^{\alpha}\dot x^{\beta}$ in the full HT metric
with an eighth-order Runge--Kutta method (relative tolerance $10^{-10}$);
the equations and initial data are given in \appref{app:orbit}.
The orbit starts on the equator at $r_{0}=17.39$~km ($x=6.13$), where the
circular orbital frequency is $\nu_{\phi}=1098.4$~Hz, equal to the upper kHz
QPO of one triplet of 4U 1728$-$34~\citep{Anand2024}.  The initial angular
velocity is that of the circular orbit, tilted by $8^{\circ}$ out of the
equatorial plane, and a radial velocity of 1~per~cent of the orbital speed is
added.  We integrate over 0.5~s of coordinate time, or 549 orbits.  Along
the orbit $|u\cdot u+1|<3\times10^{-11}$, and the energy and the axial
angular momentum are conserved to $2\times10^{-12}$ and $10^{-10}$.

\Figref[(a)]{fig:orbit} shows the first 20~ms of the orbit.  The radius
oscillates between 16.82 and 17.94~km with the radial epicyclic period
$1/\nu_{r}=2.90$~ms [panel (b)], and the particle crosses the equatorial
plane twice per vertical period $1/\nu_{\theta}=0.927$~ms, reaching
$|z|=2.5$~km [panel (c)].  The two periods differ from the orbital period
$1/\nu_{\phi}=0.910$~ms, and the orbit is therefore not closed.  At this radius
$\nu_{r}\approx\nu_{\phi}/3$, so the particle makes about three turns per radial
oscillation, and the pericentre moves forward from one radial period to the
next.  Its longitude $\varpi$ grows at the rate
$\nu_{\rm per}=\nu_{\phi}-\nu_{r}$ [panel (d)].  The line of nodes, where the orbit
crosses the equatorial plane upward, rotates in the prograde direction at
$\nu_{\rm nod}=\nu_{\phi}-\nu_{\theta}$ [panel (e)].  In 20~ms the orbital plane turns by
about $140^{\circ}$, which is visible in panel (a) as a wobbling ring.  This
is the precession of the orbital plane by frame dragging (at
$\nu_{\rm LT}=25.9$~Hz here), reduced by the torque of the stellar
oblateness ($\nu_{Q}=-4.8$~Hz in the weak-field estimate).

Measured directly from the orbit, the pericentre advances at
$\nu_{\rm per}=760.9$~Hz and the node at $\nu_{\rm nod}=19.51$~Hz.  The
epicyclic values at $r_{0}$ are 753.9 and 19.31~Hz.  The differences of 1~per~cent
are finite-amplitude effects.  For a radial velocity of 0.3~per~cent of
the orbital speed and an inclination of $0.4^{\circ}$ the same comparison at six radii between 17.5
and 47~km agrees to 0.01--0.02~per~cent, and doubling the amplitudes multiplies the
differences by four, as expected for second-order corrections.  The
epicyclic frequencies of the HT metric are therefore the precession
frequencies of actual orbits of small amplitude.

\begin{figure*}
\includegraphics[width=\textwidth]{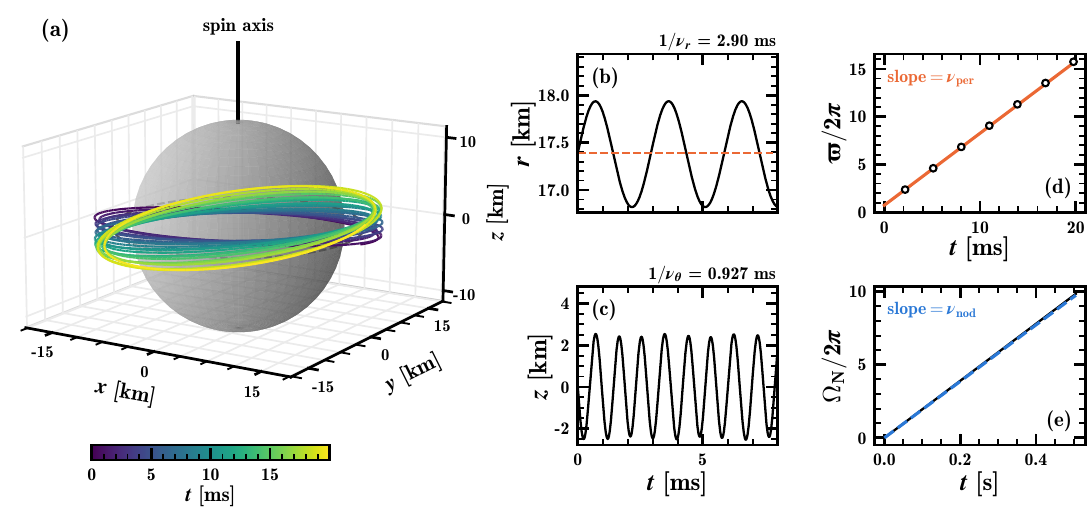}
\caption{\label{fig:orbit}Geodesic around an IOPB-I star of $1.922\,\mathrm{M}_\odot$
spinning at 363~Hz, started at $r_{0}=17.39$~km with an inclination of
$8^{\circ}$ and a radial velocity of 1~per~cent of the orbital speed.  (a) First
20~ms of the orbit in the coordinates $x=r\sin\theta\cos\phi$,
$y=r\sin\theta\sin\phi$, $z=r\cos\theta$ (true scale); the colour gives the
time.  The grey sphere has the equatorial radius of the star and the dashed
circle is the ISCO.  (b) Radial and (c) vertical oscillation; the dashed line
in (b) is $r_{0}$.  (d) Longitude of the pericentre, measured at successive
pericentres (circles), and the line of slope $\nu_{\rm per}$ predicted by the
epicyclic frequencies.  (e) Longitude of the ascending node (solid) and the
line of slope $\nu_{\rm nod}$ (dashed) over 0.5~s.}
\end{figure*}

\subsection{Light curve and power spectrum of a hot spot}

Suppose that a small, isotropically emitting spot of gas moves on this
orbit.  A distant observer in the direction $\hat n$ receives photons with
the redshift factor
\begin{equation}
g=\frac{\nu_{\rm obs}}{\nu_{\rm em}}=\frac{1}{u^{t}\left(1-\hat n\cdot\bm{v}\right)},
\label{eq:g}
\end{equation}
where $\bm{v}=d\bm{x}/dt$ is the coordinate velocity of the spot.  The factor
$u^{t}$ contains the gravitational and transverse Doppler shifts and
$1-\hat n\cdot\bm{v}$ the longitudinal Doppler shift.  Because $I_{\nu}/\nu^{3}$ is
invariant, the bolometric flux of a small spot scales as $F\propto g^{4}$.  We
neglect light bending, which changes the shape of the pulse but not its
periodicities, and include the light-travel time across the orbit,
$t_{\rm obs}=t-\hat n\cdot\bm{x}$.  The observer is inclined by $60^{\circ}$ to the
spin axis.  \appref{app:orbit} derives the redshift factor and shows which
combinations of the orbital frequencies the flux contains.

The light curve [\figref[(a)]{fig:psd}] is dominated by Doppler boosting:
the flux rises to 4.5 times its mean when the spot approaches the observer
and falls to 0.18 times the mean when it recedes, once per orbit.  The power
spectrum of 0.5~s of data [\figref[(b)]{fig:psd}] has its strongest peak at
the orbital frequency and its harmonics.  Below these it shows a series of
weaker peaks, at 19.8, 340, 762 and 1082~Hz (frequency resolution 2~Hz),
which are $\nu_{\rm nod}$, $\nu_{r}$, $\nu_{\rm per}$ and $\nu_{\theta}$ of the orbit.  Their
origin is kinematic.  The line-of-sight velocity varies at $\nu_{\phi}$, and the
radial epicycle modulates its amplitude at $\nu_{r}$, which produces sidebands at
$\nu_{\phi}\pm\nu_{r}$.  The lower sideband is the periastron precession frequency.
The vertical motion adds a term at $\nu_{\theta}$, of amplitude proportional to
$\sin\iota\cos i$, where $\iota$ is the orbital inclination and $i$ that of the
observer.  Because $F\propto g^{4}$ is a nonlinear function of the velocity,
the orbital and vertical terms beat, and the difference frequency
$\nu_{\phi}-\nu_{\theta}$ is the nodal precession frequency.  The flux of matter on a
precessing orbit therefore contains the three frequencies of the RPM.

The shaded bands in \figref[(b)]{fig:psd} are the observed triplet of
4U 1728$-$34 whose upper kHz QPO fixed the radius.  The periastron frequency
lies 2.6~per~cent (epicyclic value, 754~Hz) to 3.6~per~cent (peak, 762~Hz) above the
observed lower kHz QPO ($734.9\pm1.2$~Hz).  Part of this is the finite
amplitude of the orbit, and part reflects the fit of this star to the whole
data set (\secref{sec:obs}).  The
observed low-frequency QPO ($38.6\pm1.7$~Hz) coincides with $2\nu_{\rm nod}$, as
assumed in the RPM.  The hot spot, however, puts most of its low-frequency
power at $\nu_{\rm nod}$ itself, and its peak at $2\nu_{\rm nod}$ is weaker by almost
three orders of magnitude.  The factor 2 of the RPM therefore calls for an emitting
structure that modulates the flux twice per precession cycle, as assumed by
\citet{StellaVietriMorsink1999}, rather than a single
orbiting spot.

The amplitudes of the peaks also show the limits of the picture.  Relative to
the orbital peak the precession peaks are $10^{-3}$--$10^{-2}$ in power, whereas
the observed lower and upper kHz QPOs have comparable amplitudes.  The hot
spot shows which frequencies an orbit can imprint on the flux, not how
strongly a real accretion flow does so.

\begin{figure*}
\includegraphics[width=\textwidth]{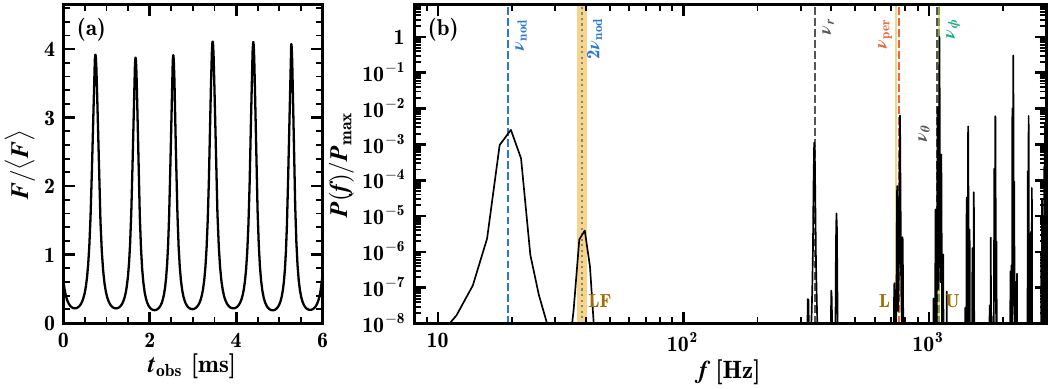}
\caption{\label{fig:psd}Hot spot on the orbit of \figref{fig:orbit}, seen at
an inclination of $60^{\circ}$.  (a) Bolometric flux, $F\propto g^{4}$, over the first
6~ms.  (b) Power spectrum of 0.5~s of the light curve, normalised to its
maximum.  Dashed lines are the epicyclic frequencies at $r_{0}$:
$\nu_{\rm nod}$, $\nu_{r}$, $\nu_{\rm per}$, $\nu_{\theta}$ and $\nu_{\phi}$; the dotted line is
$2\nu_{\rm nod}$.  The shaded bands are the low-frequency QPO (LF) and the lower
(L) and upper (U) kHz QPOs of the 4U 1728$-$34 triplet with
$\nu_{U}=1098.4$~Hz measured by \citet{Anand2024} (widths:
$1\sigma$ errors, at least 1.2~per~cent).}
\end{figure*}

\section{Masses and spins from the observed QPOs}
\label{sec:obs}

\subsection{Data and fitting}

We use four published data sets (\tabref{tab:data}).  The atoll
sources have spin frequencies known from burst oscillations, 581~Hz for
4U 1636$-$53~\citep{StrohmayerMarkwardt2002} and 363~Hz for
4U 1728$-$34~\citep{Strohmayer1996}.  For 4U 1636$-$53 we use the 28 pairs of
twin kHz QPOs of \citet{Zhang2017} from RXTE data, as tabulated by
\citet{Shi2024}.  For 4U 1728$-$34 we use the 16 sets of
simultaneous QPOs detected by \citet{Anand2024} in AstroSat/LAXPC
data; 13 are complete triplets of a low-frequency QPO and twin kHz QPOs.
The lower kHz QPO is measured to about 1~Hz, which makes this the most
precise data set.  The Z sources have simultaneous HBOs and twin kHz QPOs
but unknown spins.  For GX 5$-$1 we pair the HBO and kHz QPO frequencies of
the same selections on the Z track measured by \citet{Jonker2002}, and for GX 340+0 we use the nine selections of \citet{Jonker2000} in which both kHz peaks have free frequencies.
Asymmetric errors were symmetrised.

\begin{table}
\caption{\label{tab:data}QPO data.  $N$ is the number of QPO sets; ranges are
in Hz.  The low-frequency QPO of 4U 1728$-$34 and the HBOs of the Z sources
are listed under $\nu_{\rm LF}$.  Sources of the data:
4U 1636$-$53~\citep{Zhang2017,Shi2024}, 4U 1728$-$34~\citep{Anand2024},
GX 5$-$1~\citep{Jonker2002} and GX 340+0~\citep{Jonker2000}; spin
frequencies from \citet{StrohmayerMarkwardt2002,Strohmayer1996}.}
\setlength{\tabcolsep}{3.5pt}
\begin{tabular}{lcccccc}
\hline
Source & Type & $\nu_s$ & $N$ & $\nu_{L}$ & $\nu_{U}$ & $\nu_{\rm LF}$\\
\hline
4U 1636$-$53 & atoll & 581 & 28 & 558--949 & 868--1223 & --\\
4U 1728$-$34 & atoll & 363 & 16 & 674--853 & 1036--1138 & 34.6--44.6\\
GX 5$-$1     & Z     & --  & 10 & 156--535 & 478--828 & 14.3--47.1\\
GX 340+0     & Z     & --  & 9  & 197--565 & 535--840 & 20.6--48.2\\
\hline
\end{tabular}
\end{table}

\begin{figure*}
\includegraphics[width=\textwidth]{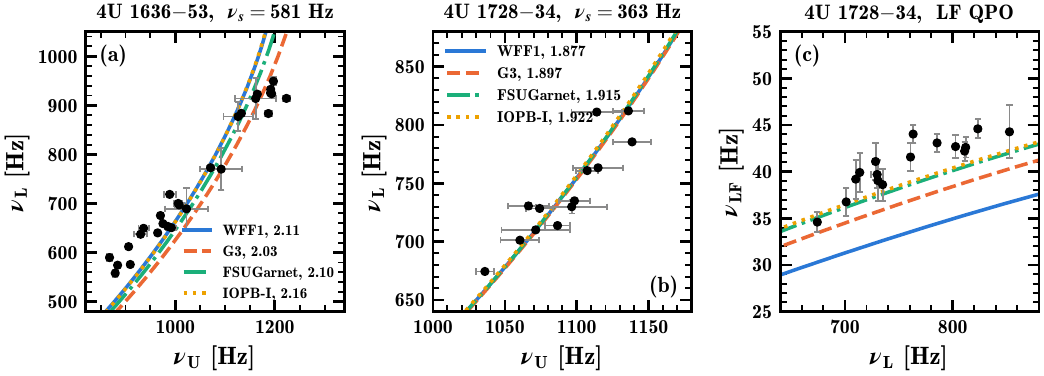}
\caption{\label{fig:tracks_atoll}Atoll sources: data and best-fitting RPM
curves for the four EOSs (best-fit masses in $\mathrm{M}_\odot$ in the legends).
(a) 4U 1636$-$53, kHz QPOs of \citet{Zhang2017} as tabulated by
\citet{Shi2024}.  (b) 4U 1728$-$34, kHz QPOs, and (c) its
low-frequency QPO against the lower kHz QPO, from AstroSat/LAXPC data of
\citet{Anand2024}.  Error bars are $1\sigma$ measurement errors.}
\end{figure*}

For a given EOS, mass and spin frequency the star fixes $(j,q,R_{\rm eq})$, and
the RPM gives $\nu_{L}$ and $\nu_{\rm LF}$ as functions of $\nu_{U}$
[\Eqref{eq:rpm}], parametrised by the radius between the inner edge of
the disc and $25M$.  The radius of each observed set of QPOs is unknown, so
we marginalise over it with a uniform weight, which can be done exactly
(\appref{app:like}).  We use Gaussian errors, to which we add in quadrature an intrinsic scatter
proportional to the frequency, one for the kHz QPOs and one for the
low-frequency QPO, because the data scatter about any RPM curve by more than
their errors.  The scatters are free parameters.  The mass has a uniform
prior on 1.0--$2.6\,\mathrm{M}_\odot$.  The spin frequency is fixed for the atoll
sources and uniform on 200--750~Hz for the Z sources.  To fit without an
EOS we replace the star by free parameters $(M,j,\bar Q)$ with uniform priors.
The posteriors are computed on grids and, for the fits of 4U 1728$-$34 and
the Z sources without an EOS, by nested
sampling~\citep{Ashton2019,Speagle2020}.  \tabref{tab:fits} lists the
results, and \figref{fig:tracks_atoll} compares the best-fitting curves
with the data of the atoll sources.

\begin{table*}
\caption{\label{tab:fits}Masses and spins from the QPOs (medians and 68~per~cent intervals).  For the atoll sources $\nu_s$ is fixed by the burst oscillations; for the Z sources it is a free parameter.  $j$, $\bar Q=-q/j^{2}$ and $I$ follow from $M$, $\nu_s$ and the EOS.  4U 1636$-$53: kHz QPOs only; the other sources: kHz QPOs and the low-frequency QPO or HBO, identified with $2\nu_{\rm nod}$.}
\begin{tabular}{llccccc}
\hline
Source & EOS & $M\,[\mathrm{M}_\odot]$ & $\nu_s$ [Hz] & $j$ & $\bar Q$ & $I\,[10^{45}\,{\rm g\,cm^{2}}]$ \\
\hline
4U 1636$-$53 & WFF1 & $2.105^{+0.025}_{-0.028}$ & 581 & $0.191^{+0.003}_{-0.003}$ & $1.58^{+0.08}_{-0.08}$ & $2.04^{+0.02}_{-0.03}$ \\
 & G3 & $2.023^{+0.005}_{-0.012}$ & 581 & $0.213^{+0.005}_{-0.004}$ & $2.13^{+0.12}_{-0.08}$ & $2.10^{+0.03}_{-0.04}$ \\
 & FSUGarnet & $2.096^{+0.008}_{-0.016}$ & 581 & $0.236^{+0.006}_{-0.004}$ & $2.42^{+0.14}_{-0.10}$ & $2.49^{+0.02}_{-0.03}$ \\
 & IOPB-I & $2.154^{+0.023}_{-0.029}$ & 581 & $0.246^{+0.006}_{-0.006}$ & $2.51^{+0.17}_{-0.17}$ & $2.74^{+0.01}_{-0.02}$ \\
\hline
4U 1728$-$34 & WFF1 & $1.877^{+0.010}_{-0.010}$ & 363 & $0.1308^{+0.0004}_{-0.0004}$ & $2.16^{+0.03}_{-0.03}$ & $1.78^{+0.01}_{-0.01}$ \\
 & G3 & $1.897^{+0.009}_{-0.010}$ & 363 & $0.153^{+0.001}_{-0.001}$ & $2.78^{+0.05}_{-0.05}$ & $2.12^{+0.01}_{-0.01}$ \\
 & FSUGarnet & $1.916^{+0.010}_{-0.010}$ & 363 & $0.171^{+0.001}_{-0.001}$ & $3.28^{+0.05}_{-0.05}$ & $2.42^{+0.01}_{-0.01}$ \\
 & IOPB-I & $1.922^{+0.010}_{-0.010}$ & 363 & $0.178^{+0.001}_{-0.001}$ & $3.47^{+0.05}_{-0.05}$ & $2.53^{+0.01}_{-0.02}$ \\
\hline
GX 5$-$1 & WFF1 & $2.127^{+0.028}_{-0.042}$ & $744^{+4}_{-11}$ & $0.240^{+0.005}_{-0.006}$ & $1.57^{+0.13}_{-0.10}$ & $2.04^{+0.01}_{-0.04}$ \\
 & G3 & $2.031^{+0.011}_{-0.021}$ & $743^{+5}_{-12}$ & $0.274^{+0.007}_{-0.007}$ & $2.31^{+0.16}_{-0.13}$ & $2.14^{+0.01}_{-0.02}$ \\
 & FSUGarnet & $2.123^{+0.007}_{-0.012}$ & $740^{+7}_{-12}$ & $0.292^{+0.006}_{-0.005}$ & $2.47^{+0.12}_{-0.08}$ & $2.50^{+0.02}_{-0.02}$ \\
 & IOPB-I & $2.206^{+0.008}_{-0.014}$ & $735^{+10}_{-13}$ & $0.294^{+0.007}_{-0.008}$ & $2.36^{+0.13}_{-0.09}$ & $2.73^{+0.01}_{-0.02}$ \\
\hline
GX 340+0 & WFF1 & $2.122^{+0.031}_{-0.047}$ & $743^{+5}_{-12}$ & $0.240^{+0.005}_{-0.006}$ & $1.57^{+0.15}_{-0.11}$ & $2.04^{+0.01}_{-0.05}$ \\
 & G3 & $2.027^{+0.013}_{-0.023}$ & $741^{+7}_{-14}$ & $0.274^{+0.008}_{-0.007}$ & $2.33^{+0.19}_{-0.13}$ & $2.14^{+0.01}_{-0.02}$ \\
 & FSUGarnet & $2.118^{+0.009}_{-0.016}$ & $738^{+9}_{-15}$ & $0.294^{+0.007}_{-0.008}$ & $2.52^{+0.15}_{-0.12}$ & $2.51^{+0.01}_{-0.03}$ \\
 & IOPB-I & $2.194^{+0.014}_{-0.021}$ & $731^{+13}_{-19}$ & $0.297^{+0.009}_{-0.011}$ & $2.46^{+0.16}_{-0.14}$ & $2.74^{+0.01}_{-0.02}$ \\
\hline
\end{tabular}
\end{table*}

\subsection{\texorpdfstring{4U 1636$-$53}{4U 1636-53}}

The twin kHz QPOs can be described with every EOS, but
only with an intrinsic scatter of 4--5~per~cent, or 40--50~Hz at
$\nu_{U}\approx1000$~Hz: the data scatter about every RPM curve
[\figref[(a)]{fig:tracks_atoll}].  The masses are high, 2.02--$2.15\,\mathrm{M}_\odot$,
and lie within $0.06\,\mathrm{M}_\odot$ of the maximum masses of the stars at 581~Hz
(2.03, 2.11, 2.16 and $2.19\,\mathrm{M}_\odot$ for G3, FSUGarnet, WFF1 and IOPB-I).  Two
effects combine.  The RPM mass of this source is known to be close to
$2\,\mathrm{M}_\odot$~\citep{Torok2012}.  In addition, at 581~Hz the spin of the star is
large, $j=0.19$--0.25, and the mass inferred from kHz QPOs grows with $j$:
a fit without an EOS prefers $M=1.89^{+0.08}_{-0.05}\,\mathrm{M}_\odot$ with a small spin,
$j=0.05^{+0.07}_{-0.04}$.  Given the large scatter, these data do not
discriminate between the EOSs.

\subsection{\texorpdfstring{4U 1728$-$34}{4U 1728-34}}

The kHz QPOs alone are fitted with a scatter below
0.6~per~cent and give $M=1.88$--$1.92\,\mathrm{M}_\odot$, depending on the EOS.  They fix the
mass for a given $j$, but their evidence is the same for all four EOSs to
within $\Delta\ln Z=0.4$.  The low-frequency QPO,
identified with $2\nu_{\rm nod}$, separates them
[\figref[(c)]{fig:tracks_atoll}].  At the fitted masses
($M=1.877$--$1.922\,\mathrm{M}_\odot$) the four EOSs give
$I=1.78$, 2.12, 2.42 and $2.53\times10^{45}$~g\,cm$^{2}$ (WFF1, G3, FSUGarnet,
IOPB-I), and the nodal frequency grows with $I$ through $j=2\pi\nu_sI/M^{2}$.
IOPB-I and FSUGarnet reproduce the low-frequency QPO with a scatter of 5--6~per~cent,
comparable to its measurement errors, while G3 and WFF1 predict nodal
frequencies that are systematically too low and need 10~per~cent and 17~per~cent.  The
Bayesian evidence accordingly favours IOPB-I and FSUGarnet, by
$\Delta\ln Z\approx7$--8 over G3 and 16--18 over WFF1; the difference between
IOPB-I and FSUGarnet ($\Delta\ln Z=1.3$) is not significant.  The preferred star has $M=1.922\pm0.010\,\mathrm{M}_\odot$, $j=0.178$ and
$\bar Q=3.47$.

\begin{figure*}
\includegraphics[width=0.72\textwidth]{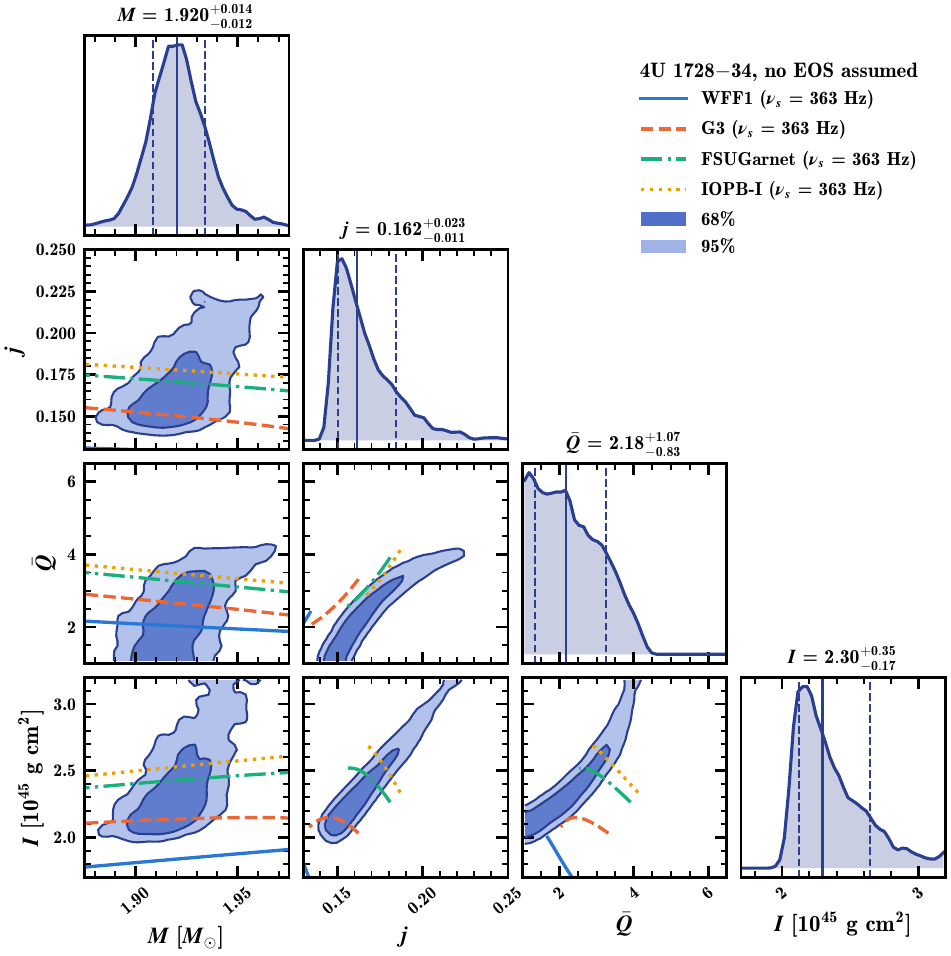}
\caption{\label{fig:corner}Posterior of the fit to 4U 1728$-$34 without an
EOS: mass, spin, reduced quadrupole and the moment of inertia implied by
the spin frequency of 363~Hz.  Shaded regions contain 68~per~cent and 95~per~cent of the
posterior; the titles give medians and 68~per~cent intervals.  Lines are the
neutron-star sequences of the four EOSs at 363~Hz.}
\end{figure*}

\Figref{fig:corner} shows the fit without an EOS.  The QPOs give
$M=1.920^{+0.014}_{-0.012}\,\mathrm{M}_\odot$, $j=0.162^{+0.023}_{-0.011}$ and $\bar Q=2.2^{+1.1}_{-0.8}$, and with the known spin
frequency $I=2.30^{+0.35}_{-0.17}\times10^{45}$~g\,cm$^{2}$.  The mass agrees with the
EOS fits.  The spin and the quadrupole are strongly correlated (correlation coefficient
0.94): the posterior is a narrow ridge along which a larger quadrupole
cancels part of a larger Lense--Thirring precession [\Eqref{eq:ratio}].
The one-dimensional ranges of $j$ and $I$ contain the stiff EOSs, but the
neutron-star sequences lie above the ridge.  At $j=0.171$ and 0.178 the QPOs
prefer $\bar Q=2.7\pm0.15$ and $3.0\pm0.13$, while FSUGarnet and IOPB-I have
$\bar Q=3.28$ and 3.47.  A neutron star with the required moment of inertia is
thus 15--20~per~cent more oblate than the QPOs prefer.  For G3 and WFF1 the offset
is larger.  The EOS fits absorb this tension in the 5--6~per~cent scatter of the
low-frequency QPO, which the fit without an EOS does not need (0.7~per~cent).

The quadrupole matters for this inference.  At the radius where
$\nu_{\phi}=1098.4$~Hz, the preferred IOPB-I star has $\nu_{\rm nod}=19.31$~Hz.  A Kerr
spacetime with the same mass and the same $j$ would give 24.6~Hz.  To
produce 19.31~Hz the Kerr spacetime needs $j=0.138$, 22~per~cent less, and with the
known spin frequency this corresponds to $I=1.97\times10^{45}$~g\,cm$^{2}$
instead of 2.53.  The Kerr value is close to the
$I\approx2\times10^{45}$~g\,cm$^{2}$ that \citet{Anand2024} obtained
from these data in the Kerr metric.  Their estimate that a self-consistent
metric would change $I$ by about 5~per~cent is, however, much smaller than the
difference found here.  An analysis in the Kerr metric would therefore select a
softer EOS, close to G3.  The size of this bias is set by the ratio of
\Eqref{eq:ratio} near the ISCO, that is, by $j$ and $\bar Q$ of the star,
and not by details of the data.

\subsection{Z sources}

\begin{figure*}
\includegraphics[width=0.92\textwidth]{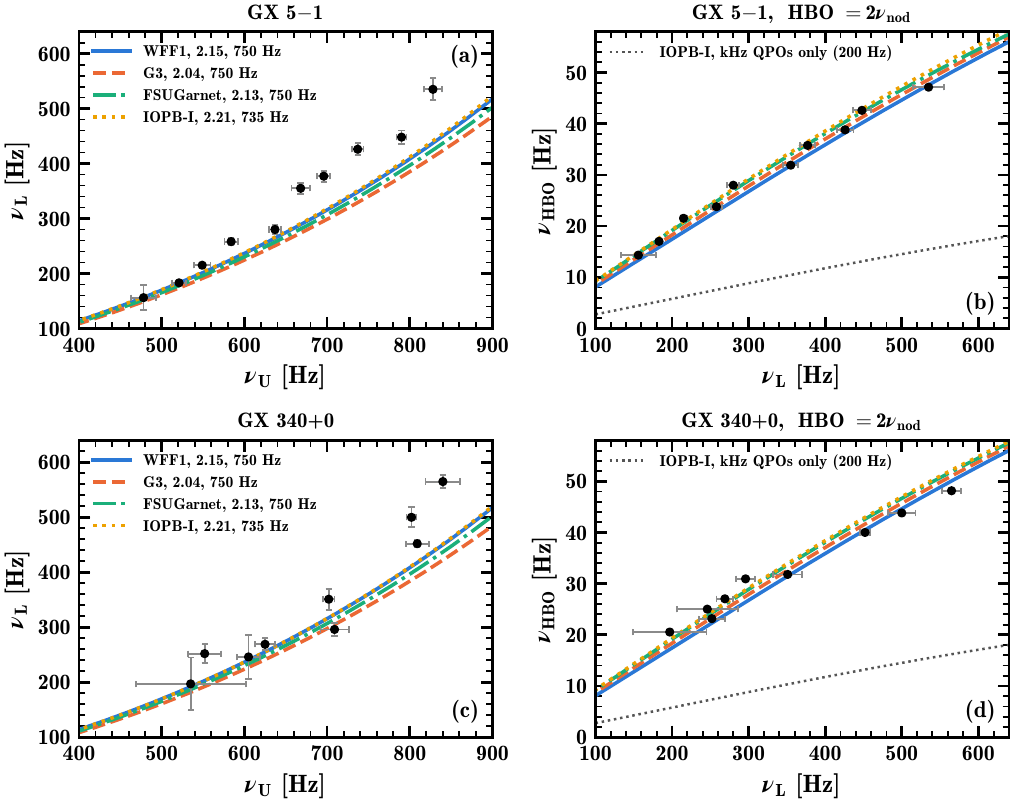}
\caption{\label{fig:tracks_z}Z sources with the best-fitting RPM curves for
the four EOSs (best-fit mass in $\mathrm{M}_\odot$ and spin frequency in the legends).
(a), (b) GX 5$-$1, data of \citet{Jonker2002}; (c), (d) GX 340+0,
data of \citet{Jonker2000}.  Panels (a) and (c) show the kHz
QPOs, panels (b) and (d) the HBOs.  The dotted grey line is $2\nu_{\rm nod}$
predicted by the fit of IOPB-I to the kHz QPOs alone.}
\end{figure*}

We first fit the kHz QPOs alone.  In GX 5$-$1 and GX 340+0 they extend to
much lower frequencies than in the atoll sources ($\nu_{L}\approx160$--200~Hz at
$\nu_{U}\approx480$--540~Hz).  In the RPM a ratio $\nu_{L}/\nu_{U}\approx0.33$ places the
emission near $x\approx10$, and the observed $\nu_{U}$ then requires a high mass.
The kHz-only fits give $M=1.97$--$2.14\,\mathrm{M}_\odot$ with a scatter of 3--7~per~cent and do
not constrain the spin.  The predicted nodal frequency is far below the
HBO: for IOPB-I, $2\nu_{\rm nod}=6$, 12 and 15~Hz at $\nu_{L}=200$, 400 and 500~Hz,
against observed HBOs of 18--48~Hz (dotted lines in \figref[(b)]{fig:tracks_z} and~\hyperref[fig:tracks_z]{\ref*{fig:tracks_z}(d)}).

We then add the HBO.  To raise $2\nu_{\rm nod}$ to the observed HBO
frequencies the fits push the spin to the upper edge of the prior,
$\nu_s=731$--744~Hz for all EOSs, and the mass to within
0.01--$0.05\,\mathrm{M}_\odot$ of the maximum mass at that spin
(\tabref{tab:fits}).  The HBO is then reproduced, but the kHz QPOs are
not.  At $\nu_{U}\gtrsim700$~Hz the observed $\nu_{L}$ lies up to 100~Hz above every
curve, and the kHz scatter grows to 5--9~per~cent.  IOPB-I and FSUGarnet, the stars
with the largest moments of inertia, come closest.

The reason is the saturation of the nodal frequency shown in
\figref[(b)]{fig:nodal}.  Raising the spin increases the Lense--Thirring
term, but the quadrupole term grows faster.  For EOSs with
$M_{\max}\approx2.0$--$2.2\,\mathrm{M}_\odot$ no spin below 750~Hz produces nodal frequencies as
high as the HBOs at the radii fixed by the kHz QPOs.  The fit without an
EOS confirms this.  It requires $M=2.49\pm0.05\,\mathrm{M}_\odot$ and $j=0.27\pm0.02$ for
GX 5$-$1 and $M=2.47^{+0.06}_{-0.07}\,\mathrm{M}_\odot$ and $j=0.34\pm0.03$ for GX 340+0, with a
kHz scatter of only 3~per~cent.  These masses exceed the maximum mass of every EOS
in our set and the masses of the heaviest known neutron
stars~\citep{Fonseca2021}.  Within the RPM and the slow-rotation
approximation, the HBOs of these Z sources can be twice the nodal
precession frequency at the radius of the kHz QPOs only if the star is
heavier than any neutron star measured so far.

\section{Discussion}
\label{sec:discussion}

\subsection{What the QPOs measure}

In the RPM the kHz QPOs constrain mainly a
combination of $M$ and $j$.  For a known spin frequency the EOS fixes $j$, so
the kHz QPOs give an EOS-dependent mass.  The low-frequency QPO adds the
nodal precession, which depends on $j$ and $q$ at the same order.  Through
$j=2\pi\nu_sI/M^{2}$ and $\bar Q(I)$ it measures the moment of inertia, and the
quadrupole has to be that of the same star.  For 4U 1728$-$34 the
EOS-consistent analysis favours $I\simeq2.4$--$2.5\times10^{45}$~g\,cm$^{2}$ at
$M\simeq1.92\,\mathrm{M}_\odot$, that is, the stiffer EOSs with $R_{1.4}\simeq13$~km.  A Kerr
spacetime with the same nodal frequency gives an $I$ lower by 22~per~cent.  This
bias applies to any EOS inference that uses the RPM nodal frequency in a
Kerr spacetime or with $q$ treated as independent of $j$.  \citet{Anand2024} analysed the same data with the RPM in the Kerr metric;
if the RPM holds, moments of inertia derived from such analyses should be
revised upward by a similar amount.

\subsection{Limits of the model}

All of these statements hold within the RPM, and the
model is under strain.  The fits need intrinsic scatters below 0.6~per~cent (kHz QPOs of
4U 1728$-$34), 4--6~per~cent (4U 1636$-$53 and the low-frequency QPO of 4U 1728$-$34)
and 5--9~per~cent (Z sources).  Only the first is comparable to the measurement
errors.  For 4U 1636$-$53 the masses lie at the maximum masses of
all four EOSs.  For the Z sources no EOS-consistent star reproduces the HBOs
as nodal precession.  The preference for the stiffer EOSs in 4U 1728$-$34 is
therefore conditional on the RPM identification of the low-frequency QPO,
and it should be read as a demonstration of how much the quadrupole matters
rather than as an EOS measurement.

\subsection{Beyond test particles}

The RPM treats the QPOs as test-particle
frequencies.  \Secref{sec:orbits} shows that orbits of finite amplitude
precess at the epicyclic rates and that a hot spot on such an orbit produces peaks at these
frequencies.  It also shows two points that the model leaves open: the hot spot
puts its low-frequency power at $\nu_{\rm nod}$, not at $2\nu_{\rm nod}$, and the
precession peaks are much weaker than the orbital peak.  Finite amplitudes
shift the frequencies by about 1~per~cent for the orbit of \figref{fig:orbit}
(radial amplitude 3~per~cent, inclination $8^{\circ}$), and the shift grows as the
square of the amplitude; for somewhat larger orbits it becomes comparable to
the scatters above.  Pressure, viscosity, radiation drag and magnetic stresses
in the flow shift the frequencies further.  With $\nu_{\rm LF}=\nu_{\rm nod}$
instead of $2\nu_{\rm nod}$, the low-frequency QPOs of all sources would be even
harder to reach.

\subsection{Slow rotation and data}

The HT metric is correct to second order in the spin.
For $q=-j^{2}$ its epicyclic ratios differ from those of the exact Kerr metric
at the same orbital frequency by at most $6\times10^{-5}$ at $j=0.1$ and
$3\times10^{-4}$ at $j=0.2$ for $x\ge6$.  Taken as an estimate of the neglected
terms, this shifts $2\nu_{\rm nod}$ by about 0.1~Hz for 4U 1728$-$34, far
below the differences between the EOSs and the 1.7~Hz error of the
low-frequency QPO.  Comparisons with numerical models~\citep{Berti2005} show that the HT
approximation gives the ISCO and the orbital frequencies to about 1~per~cent for
the spin frequencies of most known millisecond pulsars.  The higher multipoles and the stellar shape lose
accuracy as the spin approaches the mass-shedding limit, so the fits of
4U 1636$-$53 at 581~Hz and especially of the Z sources at 700--750~Hz should be
repeated with numerically exact rotating
models~\citep{Stergioulas1995,MorsinkStella1999}.  The saturation of the nodal
frequency that excludes the HBOs is driven by the quadrupole growing faster
than $\nu_s$.  Higher-order rotation terms make the stars more oblate, so we do
not expect them to reverse this conclusion; we will test this with
numerically exact models in Paper~II.

The QPO frequencies were taken from published tables.  They come from two
instruments, RXTE and AstroSat, and from different analysis methods.  For
GX 5$-$1 the kHz selections overlap, so neighbouring sets are not strictly
independent, and the uncertainties of the fitted parameters of this source
are somewhat underestimated.

\section{Conclusions}
\label{sec:conclusions}

We have studied the relativistic precession model in the Hartle--Thorne
spacetimes of slowly rotating neutron stars built from four EOSs.  For each
star the mass, spin frequency and EOS fix $j$, $q$ and the inner edge of the
disc, and the orbital, epicyclic and precession frequencies follow from the
full exterior metric.  We have followed test particles on eccentric, tilted
orbits around such a star and confronted the model with the QPOs of two
atoll and two Z sources.  Our main conclusions are as follows.

The quadrupole of the star controls the nodal precession.  At
$M\simeq1.9\,\mathrm{M}_\odot$ the reduced quadrupole of the four EOSs is
$\bar Q=2.1$--3.6, so the quadrupole is 2--4 times that of a Kerr black hole
with the same $j$.  Its retrograde contribution lowers the nodal frequency
near the ISCO by 7--25~per~cent relative to the Kerr value at 363~Hz, and it grows
faster with the spin than the Lense--Thirring term.  At fixed orbital
frequency the nodal frequency therefore has a maximum as a function of the
spin, $2\nu_{\rm nod}\approx32$--34~Hz at $\nu_{U}=900$~Hz for $1.9\,\mathrm{M}_\odot$ stars of the
stiff EOSs, and a fast spin alone cannot produce a high nodal frequency.

Orbits of finite amplitude precess at the epicyclic rates, and these rates appear in
the flux.  An eccentric, tilted geodesic around a $1.92\,\mathrm{M}_\odot$ star
spinning at 363~Hz advances its pericentre at $\nu_{\phi}-\nu_{r}$ and turns its
orbital plane at $\nu_{\phi}-\nu_{\theta}$, within 1~per~cent of the epicyclic values for a
radial amplitude of 3~per~cent and an inclination of $8^{\circ}$.  The Doppler-boosted flux
of a hot spot on this orbit has peaks at $\nu_{\rm nod}$, $\nu_{r}$, $\nu_{\rm per}$,
$\nu_{\theta}$ and $\nu_{\phi}$.  Its low-frequency power, however, lies at
$\nu_{\rm nod}$, and the peak at $2\nu_{\rm nod}$, the frequency that the RPM identifies
with the low-frequency QPO, is weaker by almost three orders of magnitude.

For 4U 1728$-$34 the low-frequency QPO measures the moment of inertia.
The kHz QPOs alone give $M=1.88$--$1.92\,\mathrm{M}_\odot$ for all four EOSs.  The
low-frequency QPO then selects the stars with large moments of inertia: the
preferred IOPB-I star has $M=1.922\pm0.010\,\mathrm{M}_\odot$, $j=0.178$ and
$I=2.5\times10^{45}$~g\,cm$^{2}$, and FSUGarnet is close.  Without an EOS the QPOs
give $M=1.920^{+0.014}_{-0.012}\,\mathrm{M}_\odot$, $j=0.162^{+0.023}_{-0.011}$ and
$\bar Q=2.2^{+1.1}_{-0.8}$.  Along the narrow $j$--$\bar Q$ ridge allowed by the data
the preferred quadrupole is 15--20~per~cent smaller than that of a neutron star with
the required moment of inertia, a tension that the EOS fits absorb in a
5--6~per~cent scatter of the low-frequency QPO.

A Kerr spacetime biases the moment of inertia.  To reproduce the
same nodal frequency at the same orbital frequency, a Kerr spacetime needs a
spin 22~per~cent smaller than the neutron star, and therefore gives a moment of
inertia 22~per~cent too small ($1.97$ instead of $2.53\times10^{45}$~g\,cm$^{2}$ for
4U 1728$-$34).  This is much larger than the 5~per~cent estimated for these data by
\citet{Anand2024}, and it would move an EOS inference from the
stiff EOSs towards softer ones such as G3.

4U 1636$-$53 is described only with a large scatter.  Its kHz QPOs
require an intrinsic scatter of 4--5~per~cent and masses of 2.02--$2.15\,\mathrm{M}_\odot$, close
to the maximum masses of all four EOSs at 581~Hz, and they do not
discriminate between the EOSs.

For these EOSs the HBOs of the Z sources cannot be nodal precession.  For
GX 5$-$1 and GX 340+0 no EOS-consistent star reproduces the HBOs as twice the
nodal frequency at the radius of the kHz QPOs.  The fits are driven to spins
of 730--745~Hz and to the maximum mass and still need a kHz scatter of
5--9~per~cent.  A fit without an EOS requires $M\approx2.5\,\mathrm{M}_\odot$, above the maximum mass
of every EOS in our set.

Taken together, these results show that an EOS-consistent quadrupole has to
be included whenever the nodal precession frequency is used to infer the
spin or the moment of inertia of a neutron star.  They also show that the RPM
is under strain: it needs intrinsic scatters of several per cent, high
masses, a factor of two in the nodal frequency that a single orbiting spot
does not produce, and it fails for the HBOs of the Z sources.  The
preference for stiff EOSs in 4U 1728$-$34 is therefore conditional on the
RPM identification of the low-frequency QPO.  Three steps would sharpen
these conclusions: numerically exact rotating stars for spin frequencies
above about 500~Hz, which we will present in Paper~II; more QPO triplets with
low-frequency QPOs measured to better than 1~per~cent, which would show whether the
5~per~cent residuals are random or systematic; and joint fits with independent
mass--radius measurements of neutron stars.

\section*{Acknowledgements}

I thank Ranjeev Misra for introducing me to this problem and
for discussions on the relativistic precession model and QPOs.
I also thank H. C. Das for providing the unified equation-
of-state tables of the G3, FSUGarnet and IOPB-I models \citep{Parmar2022}, publicly available at
\url{https://github.com/hcdas/Unified_eos}, and N.~Stergioulas for making
the RNS code and its equation-of-state tables publicly
available \citep{Stergioulas1995}.  

\section*{Data Availability}

The QPO frequencies used in this work are taken from the published
literature cited in \tabref{tab:data}.  The code that computes the stars,
the frequencies, the orbits and their spectra, and the fits, together with
the QPO data, the stellar models, the frequency tables and the fit results,
is available at \url{https://github.com/AnkitKhunt/slow-rotation-NS-QPO}.

\bibliographystyle{mnras}
\bibliography{refs}

@article{Hartle1967, author={Hartle, James B.}, title={Slowly rotating relativistic stars. {I}. {E}quations of structure},
 journal={ApJ}, volume={150}, pages={1005}, year={1967}, doi={10.1086/149400}}

@article{HartleThorne1968, author={Hartle, James B. and Thorne, Kip S.}, title={Slowly rotating relativistic stars. {II}. {M}odels for neutron stars and supermassive stars},
 journal={ApJ}, volume={153}, pages={807}, year={1968}, doi={10.1086/149707}}

@article{Berti2005, author={Berti, Emanuele and White, Frances and Maniopoulou, Asimina and Bruni, Marco},
 title={Rotating neutron stars: an invariant comparison of approximate and numerical space-time models},
 journal={MNRAS}, volume={358}, pages={923}, year={2005}, doi={10.1111/j.1365-2966.2005.08812.x}}

@article{StellaVietri1998, author={Stella, Luigi and Vietri, Mario}, title={{Lense-Thirring} precession and quasi-periodic oscillations in low-mass {X}-ray binaries},
 journal={ApJ}, volume={492}, pages={L59}, year={1998}, doi={10.1086/311075}}

@article{StellaVietri1999, author={Stella, Luigi and Vietri, Mario}, title={{kHz} quasiperiodic oscillations in low-mass {X}-ray binaries as probes of general relativity in the strong-field regime},
 journal={Phys. Rev. Lett.}, volume={82}, pages={17}, year={1999}, doi={10.1103/PhysRevLett.82.17}}

@article{StellaVietriMorsink1999, author={Stella, Luigi and Vietri, Mario and Morsink, Sharon M.}, title={Correlations in the quasi-periodic oscillation frequencies of low-mass {X}-ray binaries and the relativistic precession model},
 journal={ApJ}, volume={524}, pages={L63}, year={1999}, doi={10.1086/312291}}

@article{MorsinkStella1999, author={Morsink, Sharon M. and Stella, Luigi}, title={Relativistic precession around rotating neutron stars: effects due to frame dragging and stellar oblateness},
 journal={ApJ}, volume={513}, pages={827}, year={1999}, doi={10.1086/306876}}

@article{Okazaki1987, author={Okazaki, Atsuo T. and Kato, Shoji and Fukue, Jun}, title={Global trapped oscillations of relativistic accretion disks},
 journal={PASJ}, volume={39}, pages={457}, year={1987}, doi = {10.1093/pasj/39.3.457}}

@article{vanderKlis2000, author={van der Klis, Michiel}, title={Millisecond oscillations in {X}-ray binaries},
 journal={ARA\&A}, volume={38}, pages={717}, year={2000}, doi={10.1146/annurev.astro.38.1.717}}

@incollection{vanderKlis2006, author={van der Klis, Michiel}, title={Rapid {X}-ray variability}, booktitle={Compact Stellar {X}-ray Sources}, series={Cambridge Astrophysics Series}, volume={39},
 editor={Lewin, Walter H. G. and van der Klis, Michiel}, publisher={Cambridge University Press}, address={Cambridge}, pages={39}, year={2006}, eprint={astro-ph/0410551}}

@article{PsaltisBelloniVanderKlis1999, author={Psaltis, Dimitrios and Belloni, Tomaso and van der Klis, Michiel}, title={Correlations in quasi-periodic oscillation and noise frequencies among neutron star and black hole {X}-ray binaries},
 journal={ApJ}, volume={520}, pages={262}, year={1999}, doi={10.1086/307436}}

@article{Torok2010, author={T{\"o}r{\"o}k, Gabriel and Bakala, Pavel and {\v{S}}r{\'a}mkov{\'a}, Eva and Stuchl{\'\i}k, Zden{\v{e}}k and Urbanec, Martin},
 title={On mass constraints implied by the relativistic precession model of twin-peak quasi-periodic oscillations in {Circinus X-1}},
 journal={ApJ}, volume={714}, pages={748}, year={2010}, doi={10.1088/0004-637X/714/1/748}}

@article{Torok2012, author={T{\"o}r{\"o}k, Gabriel and Bakala, Pavel and {\v{S}}r{\'a}mkov{\'a}, Eva and Stuchl{\'\i}k, Zden{\v{e}}k and Urbanec, Martin and Goluchov{\'a}, Kate{\v{r}}ina},
 title={Mass--angular-momentum relations implied by models of twin peak quasi-periodic oscillations},
 journal={ApJ}, volume={760}, pages={138}, year={2012}, doi={10.1088/0004-637X/760/2/138}, eprint={1408.4220}}

@article{Anand2024, author={Anand, Kewal and Misra, Ranjeev and Yadav, J. S. and Jain, Pankaj and Kumar, Umang and Bhattacharya, Dipankar},
 title={Detection of simultaneous quasiperiodic oscillation triplets in {4U 1728-34} and constraining the neutron star mass and moment of inertia},
 journal={ApJ}, volume={967}, pages={129}, year={2024}, doi={10.3847/1538-4357/ad410c}, eprint={2404.09546}}

@article{Anand2026, author={Anand, Kewal and Misra, Ranjeev and Yadav, J. S. and Jain, Pankaj},
 title={Correlations between {kHz QPOs} and spectral parameters from time-resolved spectro-temporal analysis of {4U 1728-34}},
 journal={ApJ}, year={2026}, doi={10.3847/1538-4357/ae53d9}, eprint={2603.16398}}

@article{Motta2014, author={Motta, S. E. and Belloni, T. M. and Stella, L. and Mu{\~n}oz-Darias, T. and Fender, R.},
 title={Precise mass and spin measurements for a stellar-mass black hole through {X}-ray timing: the case of {GRO J1655-40}},
 journal={MNRAS}, volume={437}, pages={2554}, year={2014}, doi={10.1093/mnras/stt2068}}

@article{Motta2022, author={Motta, S. E. and Belloni, T. and Stella, L. and Pappas, G. and Casares, J. and Mu{\~n}oz-Darias, T. and Torres, M. A. P. and Yanes-Rizo, I. V.},
 title={Black hole mass and spin measurements through the relativistic precession model: {XTE J1859+226}},
 journal={MNRAS}, volume={517}, pages={1469}, year={2022}, doi={10.1093/mnras/stac2142}}

@article{LaarakkersPoisson1999, author={Laarakkers, William G. and Poisson, Eric}, title={Quadrupole moments of rotating neutron stars},
 journal={ApJ}, volume={512}, pages={282}, year={1999}, doi={10.1086/306732}}

@article{Pappas2012, author={Pappas, George and Apostolatos, Theocharis A.}, title={Revising the multipole moments of numerical spacetimes and its consequences},
 journal={Phys. Rev. Lett.}, volume={108}, pages={231104}, year={2012}, doi={10.1103/PhysRevLett.108.231104}}

@article{Pappas2012MNRAS, author={Pappas, George}, title={What can quasi-periodic oscillations tell us about the structure of the corresponding compact objects?},
 journal={MNRAS}, volume={422}, pages={2581}, year={2012}, doi = {10.1111/j.1365-2966.2012.20817.x},eprint={1201.6071}}

@article{YagiYunes2013, author={Yagi, Kent and Yunes, Nicol{\'a}s}, title={{I-Love-Q}: unexpected universal relations for neutron stars and quark stars},
 journal={Science}, volume={341}, pages={365}, year={2013}, doi={10.1126/science.1236462}, eprint={1302.4499}}

@article{StrohmayerMarkwardt2002, author={Strohmayer, Tod E. and Markwardt, Craig B.}, title={Evidence for a millisecond pulsar in {4U 1636-53} during a superburst},
 journal={ApJ}, volume={577}, pages={337}, year={2002}, doi={10.1086/342152}}

@article{Strohmayer1996, author={Strohmayer, Tod E. and Zhang, William and Swank, Jean H. and others},
 title={Millisecond {X}-ray variability from an accreting neutron star system},
 journal={ApJ}, volume={469}, pages={L9}, year={1996}, doi={10.1086/310261}}

@article{Zhang2017, author={Zhang, Guobao and M{\'e}ndez, Mariano and Sanna, Andrea and Ribeiro, Evandro M. and Gelfand, Joseph D.},
 title={Relation between spectral changes and the presence of the lower {kHz QPO} in the neutron-star low-mass {X}-ray binary {4U 1636-53}},
 journal={MNRAS}, volume={465}, pages={5003}, year={2017},doi = {10.1093/mnras/stw3100}, eprint={1611.09097}}

@article{Shi2024, author={Shi, Chang-Sheng and Zhang, Guo-Bao and Zhang, Shuang-Nan and Li, Xiang-Dong},
 title={Radiation mechanism of twin kilohertz quasi-periodic oscillations in neutron star low-mass {X}-ray binaries},
 journal={A\&A}, volume={691}, pages={A187}, year={2024}, doi={10.1051/0004-6361/202449728}}

@article{Jonker2002, author={Jonker, Peter G. and van der Klis, Michiel and Homan, Jeroen and M{\'e}ndez, Mariano and Lewin, Walter H. G. and Wijnands, Rudy and Zhang, William},
 title={Low- and high-frequency variability as a function of spectral properties in the bright {X}-ray binary {GX 5-1}},
 journal={MNRAS}, volume={333}, pages={665}, year={2002}, doi={10.1046/j.1365-8711.2002.05442.x}, eprint={astro-ph/0202420}}

@article{Jonker2000, author={Jonker, Peter G. and van der Klis, Michiel and Wijnands, Rudy and Homan, Jeroen and van Paradijs, Jan and M{\'e}ndez, Mariano and Ford, Eric C. and Kuulkers, Erik and Lamb, Frederick K.},
 title={The power spectral properties of the {Z} source {GX 340+0}},
 journal={ApJ}, volume={537}, pages={374}, year={2000}, doi={10.1086/309029}, eprint={astro-ph/0002022}}

@article{Ashton2019, author={Ashton, Gregory and others}, title={{BILBY}: a user-friendly {Bayesian} inference library for gravitational-wave astronomy},
 journal={ApJS}, volume={241}, pages={27}, year={2019}, doi={10.3847/1538-4365/ab06fc}, eprint={1811.02042}}

@article{Speagle2020, author={Speagle, Joshua S.}, title={{DYNESTY}: a dynamic nested sampling package for estimating {Bayesian} posteriors and evidences},
 journal={MNRAS}, volume={493}, pages={3132}, year={2020}, doi={10.1093/mnras/staa278}, eprint={1904.02180}}

@article{WFF1988, author={Wiringa, R. B. and Fiks, V. and Fabrocini, A.}, title={Equation of state for dense nucleon matter},
 journal={Phys. Rev. C}, volume={38}, pages={1010}, year={1988}, doi={10.1103/PhysRevC.38.1010}}

@article{Kumar2017G3, author={Kumar, Bharat and Singh, S. K. and Agrawal, B. K. and Patra, S. K.}, title={New parameterization of the effective field theory motivated relativistic mean field model},
 journal={Nucl. Phys. A}, volume={966}, pages={197}, year={2017}, doi={10.1016/j.nuclphysa.2017.07.001}, eprint={1705.02621}}

@article{ChenPiekarewicz2015, author={Chen, Wei-Chia and Piekarewicz, J.}, title={Searching for isovector signatures in the neutron-rich oxygen and calcium isotopes},
 journal={Phys. Lett. B}, volume={748}, pages={284}, year={2015}, doi={10.1016/j.physletb.2015.07.020}, eprint={1412.7870}}

@article{Kumar2018IOPB, author={Kumar, Bharat and Patra, S. K. and Agrawal, B. K.}, title={New relativistic effective interaction for finite nuclei, infinite nuclear matter, and neutron stars},
 journal={Phys. Rev. C}, volume={97}, pages={045806}, year={2018}, doi={10.1103/PhysRevC.97.045806}, eprint={1711.04940}}

@article{Stergioulas1995, author={Stergioulas, Nikolaos and Friedman, John L.}, title={Comparing models of rapidly rotating relativistic stars constructed by two numerical methods},
 journal={ApJ}, volume={444}, pages={306}, year={1995}, doi={10.1086/175605}, eprint={astro-ph/9411032}}

@article{HasingerVanderKlis1989, author={Hasinger, G{\"u}nther and van der Klis, Michiel}, title={Two patterns of correlated {X}-ray timing and spectral behaviour in low-mass {X}-ray binaries},
 journal={A\&A}, volume={225}, pages={79}, year={1989}, url={https://ui.adsabs.harvard.edu/abs/1989A&A...225...79H}}

@article{vanderKlis1996, author={van der Klis, Michiel and Swank, Jean H. and Zhang, William and Jahoda, Keith and Morgan, Edward H. and Lewin, Walter H. G. and Vaughan, Brian and van Paradijs, Jan},
 title={Discovery of submillisecond quasi-periodic oscillations in the {X}-ray flux of {Scorpius X-1}},
 journal={ApJ}, volume={469}, pages={L1}, year={1996}, doi={10.1086/310251}}

@article{Berger1996, author={Berger, M. and van der Klis, M. and van Paradijs, J. and others}, title={Discovery of 800 {Hz} quasi-periodic oscillations in {4U 1608-52}},
 journal={ApJ}, volume={469}, pages={L13}, year={1996}, doi={10.1086/310256}}

@article{Chauhan2017, author={Chauhan, J. V. and Yadav, J. S. and Misra, R. and others},
 title={{AstroSat/LAXPC} detection of millisecond phenomena in {4U 1728-34}},
 journal={ApJ}, volume={841}, pages={41}, year={2017}, doi={10.3847/1538-4357/aa6d7e}, eprint={1704.04931}}

@article{AlparShaham1985, author={Alpar, M. Ali and Shaham, Jacob}, title={Is {GX 5-1} a millisecond pulsar?},
 journal={Nature}, volume={316}, pages={239}, year={1985}, doi={10.1038/316239a0}}

@article{Lamb1985, author={Lamb, Frederick K. and Shibazaki, Noriaki and Alpar, M. Ali and Shaham, Jacob}, title={Quasi-periodic oscillations in bright galactic-bulge {X}-ray sources},
 journal={Nature}, volume={317}, pages={681}, year={1985}, doi={10.1038/317681a0}}

@article{MillerLambPsaltis1998, author={Miller, M. Coleman and Lamb, Frederick K. and Psaltis, Dimitrios}, title={Sonic-point model of kilohertz quasi-periodic brightness oscillations in low-mass {X}-ray binaries},
 journal={ApJ}, volume={508}, pages={791}, year={1998}, doi={10.1086/306408}}

@article{AbramowiczKluzniak2001, author={Abramowicz, Marek A. and Klu{\'z}niak, W{\l}odzimierz}, title={A precise determination of black hole spin in {GRO J1655-40}},
 journal={A\&A}, volume={374}, pages={L19}, year={2001}, doi={10.1051/0004-6361:20010791}}

@article{Matuszkova2024, author={Matuszkov{\'a}, Monika and T{\"o}r{\"o}k, Gabriel and Klimovi{\v{c}}ov{\'a}, Kate{\v{r}}ina and Hor{\'a}k, Ji{\v{r}}{\'\i} and Straub, Odele and {\v{S}}r{\'a}mkov{\'a}, Eva and Lan{\v{c}}ov{\'a}, Debora and Urbanec, Martin and Urbancov{\'a}, Gabriela and Karas, Vladim{\'\i}r},
 title={Accretion tori around rotating neutron stars. {II}. {O}scillations and precessions},
 journal={A\&A}, volume={691}, pages={A168}, year={2024}, doi={10.1051/0004-6361/202450058}, eprint={2403.16231}}

@article{duBuisson2019, author={du Buisson, L. and Motta, S. and Fender, R.}, title={Mass and spin measurements for the neutron star {4U1608-52} through the relativistic precession model},
 journal={MNRAS}, volume={486}, pages={4485}, year={2019}, doi={10.1093/mnras/stz1160}, eprint={1905.00366}}

@article{IngramDoneFragile2009, author={Ingram, Adam and Done, Chris and Fragile, P. Chris}, title={Low-frequency quasi-periodic oscillation spectra and {Lense-Thirring} precession},
 journal={MNRAS}, volume={397}, pages={L101}, year={2009}, doi={10.1111/j.1745-3933.2009.00693.x}, eprint={0901.1238}}

@article{IngramMotta2019, author={Ingram, Adam R. and Motta, Sara E.}, title={A review of quasi-periodic oscillations from black hole {X}-ray binaries: observation and theory},
 journal={New Astron. Rev.}, volume={85}, pages={101524}, year={2019}, doi={10.1016/j.newar.2020.101524}}

@article{vanDoesburgh2017, author={van Doesburgh, Marieke and van der Klis, Michiel},
 title={Testing the relativistic precession model using low-frequency and {kHz} quasi-periodic oscillations in neutron star low-mass {X}-ray binaries with known spin},
 journal={MNRAS}, volume={465}, pages={3581}, year={2017}, doi={10.1093/mnras/stw2961}}

@article{Pachon2012, author={Pach{\'o}n, Leonardo A. and Rueda, Jorge A. and Valenzuela-Toledo, C{\'e}sar A.},
 title={On the relativistic precession and oscillation frequencies of test particles around rapidly rotating compact stars},
 journal={ApJ}, volume={756}, pages={82}, year={2012}, doi={10.1088/0004-637X/756/1/82}}

@article{Maselli2020, author={Maselli, Andrea and Pappas, George and Pani, Paolo and Gualtieri, Leonardo and Motta, Sara and Ferrari, Valeria and Stella, Luigi},
 title={A new method to constrain neutron star structure from quasi-periodic oscillations},
 journal={ApJ}, volume={899}, pages={139}, year={2020}, doi={10.3847/1538-4357/ab9ff4}, eprint={1905.05616}}

@article{Boshkayev2014, author={Boshkayev, Kuantay and Bini, Donato and Rueda, Jorge and Geralico, Andrea and Muccino, Marco and Siutsou, Ivan},
 title={What can we extract from quasiperiodic oscillations?},
 journal={Grav. Cosmol.}, volume={20}, pages={233}, year={2014}, doi={10.1134/S0202289314040033}, eprint={1412.8214}}

@article{Boshkayev2015, author={Boshkayev, Kuantay and Rueda, Jorge and Muccino, Marco},
 title={Extracting multipole moments of neutron stars from quasi-periodic oscillations in low mass {X}-ray binaries},
 journal={Astron. Rep.}, volume={59}, pages={441}, year={2015}, doi={10.1134/S1063772915060050}}

@article{Boshkayev2024, author={Boshkayev, K. and Konysbayev, T. and Kurmanov, Ye. and Muccino, M. and Quevedo, H.},
 title={Quasi-periodic oscillations in rotating and deformed spacetimes},
 journal={MNRAS}, volume={531}, pages={3876}, year={2024}, doi={10.1093/mnras/stae1388}, eprint={2312.03630}}

@article{Boshkayev2026, author={Boshkayev, K. and Konysbayev, T. and Kurmanov, Ye. and Muccino, M. and Quevedo, H.},
 title={Numerical analysis of quasiperiodic oscillations in the {Hartle-Thorne} spacetime},
 journal={MNRAS}, volume={549}, year={2026}, doi={10.1093/mnras/stag456}, eprint={2506.11581}}

@article{Urbancova2019, author={Urbancov{\'a}, Gabriela and Urbanec, Martin and T{\"o}r{\"o}k, Gabriel and Stuchl{\'\i}k, Zden{\v{e}}k and Blaschke, Martin and Miller, John C.},
 title={Epicyclic oscillations in the {Hartle-Thorne} external geometry},
 journal={ApJ}, volume={877}, pages={66}, year={2019}, doi={10.3847/1538-4357/ab1b4c}, eprint={1905.00730}}

@article{Torok2025, author={T{\"o}r{\"o}k, Gabriel and Urbanec, Martin and Matuszkov{\'a}, Monika and Urbancov{\'a}, Gabriela and Klimovi{\v{c}}ov{\'a}, Kate{\v{r}}ina and Lan{\v{c}}ov{\'a}, Debora and others},
 title={{Lense-Thirring} precession of neutron-star accretion flows: relativistic versus classical precession},
 journal={A\&A}, volume={703}, pages={A30}, year={2025}, doi={10.1051/0004-6361/202554318}, eprint={2508.13777}}

@article{Fonseca2021, author={Fonseca, E. and others}, title={Refined mass and geometric measurements of the high-mass {PSR J0740+6620}},
 journal={ApJ}, volume={915}, pages={L12}, year={2021}, doi={10.3847/2041-8213/ac03b8}}

@article{Riley2021, author={Riley, Thomas E. and Watts, Anna L. and Ray, Paul S. and others},
 title={A {NICER} view of the massive pulsar {PSR J0740+6620} informed by radio timing and {XMM-Newton} spectroscopy},
 journal={ApJ}, volume={918}, pages={L27}, year={2021}, doi={10.3847/2041-8213/ac0a81}}

@article{Miller2021, author={Miller, M. C. and others}, title={The radius of {PSR J0740+6620} from {NICER} and {XMM-Newton} data},
 journal={ApJ}, volume={918}, pages={L28}, year={2021}, doi={10.3847/2041-8213/ac089b}, eprint={2105.06979}}

@article{Parmar2022, author={Parmar, Vishal and Das, H. C. and Kumar, Ankit and Sharma, M. K. and Patra, S. K.},
 title={Crustal properties of a neutron star within an effective relativistic mean-field model},
 journal={Phys. Rev. D}, volume={105}, pages={043017}, year={2022}, doi={10.1103/PhysRevD.105.043017}, eprint={2111.07278}}

\appendix
\section{Rotating stars to second order in the angular velocity}
\label{app:interior}

We use units with $G=c=1$ and follow \citet{Hartle1967}.  Throughout this
appendix $\nu$ denotes the metric function with $e^{\nu}=-g_{tt}$ of the
nonrotating star, $P_{2}=P_{2}(\cos\theta)=\tfrac{1}{2}(3\cos^{2}\theta-1)$, and a
prime denotes $d/dr$.

\subsection{Metric, fluid and ordering}

The metric of a rigidly rotating star is written in the form of
\Eqref{eq:htmetric}.  The rotation is slow in the sense that
$\Omega^{2}R^{3}/M\ll1$; we keep terms to second order in $\Omega$.  The
frame-dragging angular velocity $\omega(r)$ is of first order, and the
functions $h_{0}$, $h_{2}$, $m_{0}$, $m_{2}$ and $k_{2}$ are of second order.
Terms of first order would change sign with $\Omega$ and cannot appear in the
diagonal components, and the $\theta$ dependence at second order is that of
$P_{0}$ and $P_{2}$ only, because the centrifugal potential is
$\propto r^{2}\sin^{2}\theta=\tfrac{2}{3}r^{2}(1-P_{2})$.

The fluid is a perfect fluid with energy density $\varepsilon$ and pressure $P$,
$T^{\mu\nu}=(\varepsilon+P)u^{\mu}u^{\nu}+Pg^{\mu\nu}$, rotating rigidly,
$u^{\mu}=u^{t}(1,0,0,\Omega)$.  To second order,
\begin{equation}
g_{tt}+2\Omega g_{t\phi}+\Omega^{2}g_{\phi\phi}
=-e^{\nu}\left(1+2h_{0}+2h_{2}P_{2}\right)+r^{2}\sin^{2}\theta\,\varpi^{2},
\end{equation}
with $\varpi=\Omega-\omega$ the angular velocity of the fluid relative to the
local inertial frames.  The normalisation $u\cdot u=-1$ then gives
\begin{equation}
u^{t}=e^{-\nu/2}\left[1-h_{0}-h_{2}P_{2}+\tfrac{1}{2}e^{-\nu}r^{2}\sin^{2}\theta\,\varpi^{2}\right].
\label{eq:ut}
\end{equation}
In the frame comoving with the fluid the pressure at a given coordinate
point changes by $(\varepsilon+P)\,p^{*}$ with $p^{*}=p_{0}^{*}+p_{2}^{*}P_{2}$, and the
energy density by $(d\varepsilon/dP)(\varepsilon+P)\,p^{*}$, since the EOS is
barotropic.

\subsection{Nonrotating background}

At zeroth order the field equations reduce to the TOV equations,
\begin{align}
m'&=4\pi r^{2}\varepsilon,\qquad
P'=-(\varepsilon+P)\frac{m+4\pi r^{3}P}{r(r-2m)},\nonumber\\
\nu'&=\frac{2(m+4\pi r^{3}P)}{r(r-2m)},\qquad e^{\lambda}=\left(1-\frac{2m}{r}\right)^{-1}.
\label{eq:tov}
\end{align}
They are integrated from a regular centre, $m=0$, $P=P_{c}$, to the surface
$r=R$ where $P=0$.  The constant in $\nu$ is fixed by $e^{\nu(R)}=1-2M_{\rm
TOV}/R$, which matches the interior to the exterior Schwarzschild metric.
Two combinations will be used repeatedly.  Adding the $tt$ and $rr$ field
equations gives
\begin{equation}
\nu'+\lambda'=8\pi r(\varepsilon+P)e^{\lambda},
\label{eq:nulam}
\end{equation}
and near the centre, where $m\simeq\tfrac{4}{3}\pi\varepsilon_{c}r^{3}$,
\begin{equation}
\nu'\simeq\frac{8\pi}{3}\left(\varepsilon_{c}+3P_{c}\right)r .
\label{eq:nucentre}
\end{equation}

\subsection{Hydrostatic equilibrium of the rotating fluid}

For a stationary flow along the Killing vector $\xi=\partial_{t}+\Omega\,\partial_{\phi}$,
$u^{\mu}=u^{t}\xi^{\mu}$ with $u^{t}=(-\xi\cdot\xi)^{-1/2}$, and the four-acceleration
is $a_{\mu}=-\partial_{\mu}\ln u^{t}$.  The relativistic Euler equation,
$(\varepsilon+P)a_{\mu}=-\partial_{\mu}P$ for the spatial components, becomes
\begin{equation}
\frac{\partial_{i}P}{\varepsilon+P}=\partial_{i}\ln u^{t},
\end{equation}
so that for a barotropic fluid
\begin{equation}
\int^{P}\frac{dP}{\varepsilon+P}-\ln u^{t}={\rm const}.
\label{eq:firstint}
\end{equation}
At zeroth order \Eqref{eq:firstint} is the TOV equation for $P'$ written as
$P'/(\varepsilon+P)=-\nu'/2$.  At second order the first term changes by
$\delta P/(\varepsilon+P)=p^{*}$, and the second by $\delta\ln u^{t}$ from
\Eqref{eq:ut}.  With $\sin^{2}\theta=\tfrac{2}{3}(1-P_{2})$ the $P_{0}$ and $P_{2}$ parts
of \Eqref{eq:firstint} give
\begin{align}
p_{0}^{*}&=-h_{0}+\tfrac{1}{3}e^{-\nu}r^{2}\varpi^{2}+h_{0c},\label{eq:p0}\\
p_{2}^{*}&=-h_{2}-\tfrac{1}{3}e^{-\nu}r^{2}\varpi^{2},\label{eq:p2}
\end{align}
where $h_{0c}$ is a constant.  These two relations replace the
perturbed Euler equations in what follows.

\subsection{Frame dragging}

At first order in $\Omega$ the only nontrivial field equation is the $t\phi$
component.  Its source is
$T^{t}{}_{\phi}=(\varepsilon+P)u^{t}u_{\phi}=(\varepsilon+P)e^{-\nu}r^{2}\sin^{2}\theta\,\varpi$.  After
the angular dependence is separated, regularity at the centre and
asymptotic flatness leave only the $\ell=1$ part, so that $\omega=\omega(r)$,
and the equation is \citep{Hartle1967}
\begin{equation}
\frac{1}{r^{4}}\frac{d}{dr}\left(r^{4}\bar\jmath\,\frac{d\varpi}{dr}\right)+\frac{4}{r}\frac{d\bar\jmath}{dr}\,\varpi=0,
\qquad \bar\jmath=e^{-(\nu+\lambda)/2}.
\label{eq:fd}
\end{equation}
By \Eqref{eq:nulam},
\begin{equation}
\frac{d\bar\jmath}{dr}=-\tfrac{1}{2}(\nu'+\lambda')\bar\jmath=-4\pi r(\varepsilon+P)e^{\lambda}\bar\jmath ,
\end{equation}
so the second term of \Eqref{eq:fd} equals
$-16\pi(\varepsilon+P)e^{(\lambda-\nu)/2}\varpi$.  It is proportional to the angular
momentum density of the fluid and vanishes outside the star.

Near the centre, $\bar\jmath\simeq\bar\jmath_{c}$ and $e^{\lambda}\simeq1$, and
\Eqref{eq:fd} reads $\varpi''+4\varpi'/r=16\pi(\varepsilon_{c}+P_{c})\varpi$.  The regular
solution is
\begin{equation}
\varpi=\varpi_{c}\left[1+\tfrac{8\pi}{5}(\varepsilon_{c}+P_{c})r^{2}+\mathcal{O}(r^{4})\right],
\end{equation}
with $\varpi_{c}$ arbitrary.  Because \Eqref{eq:fd} is linear and homogeneous,
we integrate it with $\varpi_{c}=1$ and rescale the solution at the end.

Outside the star $\bar\jmath=1$ and $(r^{4}\varpi')'=0$, so
$\varpi=C_{1}-C_{2}/r^{3}$.  Asymptotic flatness requires $\omega\to0$, hence
$C_{1}=\Omega$, and comparison of $g_{t\phi}=-\omega r^{2}\sin^{2}\theta$ with the
gravitomagnetic field of a body of angular momentum $J$,
$g_{t\phi}\to-2J\sin^{2}\theta/r$, gives $C_{2}=2J$:
\begin{equation}
\omega=\frac{2J}{r^{3}},\qquad \varpi=\Omega-\frac{2J}{r^{3}}\qquad(r>R).
\end{equation}

Since $\varepsilon$ and $P$ are finite at the surface, $\varpi$
and $\varpi'$ are continuous there.  The exterior solution then gives
\begin{equation}
J=\tfrac{1}{6}R^{4}\varpi'(R),\qquad
\Omega=\varpi(R)+\tfrac{1}{3}R\,\varpi'(R),\qquad I=\frac{J}{\Omega}.
\label{eq:Jmatch}
\end{equation}
The second relation fixes the scale of the interior solution for a given
$\Omega$.  Integrating \Eqref{eq:fd} from the centre to $R$ gives the
equivalent volume integral
\begin{equation}
J=\frac{8\pi}{3}\int_{0}^{R}r^{4}(\varepsilon+P)\,e^{(\lambda-\nu)/2}\,\varpi\,dr ,
\label{eq:Jint}
\end{equation}
which reduces to the Newtonian $J=\Omega\,\tfrac{8\pi}{3}\int\rho r^{4}dr$ for
weak fields.  For the models of this paper equations
(\ref{eq:Jmatch}) and (\ref{eq:Jint}) agree to better than $10^{-6}$.  Since
$\varpi\propto\Omega$ at fixed central density, $J\propto\Omega$ and $I$ is
independent of $\Omega$ at this order.

\subsection{Monopole deformation}

The $P_{0}$ parts of the $tt$ and $rr$ field equations at second order
involve $m_{0}$, $h_{0}$ and $p_{0}^{*}$.  Eliminating $h_{0}$ with
\Eqref{eq:p0}, \citet{Hartle1967} obtained
\begin{align}
\frac{dm_{0}}{dr}=&\,4\pi r^{2}\frac{d\varepsilon}{dP}(\varepsilon+P)\,p_{0}^{*}
+\frac{1}{12}\bar\jmath^{2}r^{4}\varpi'^{2}-\frac{1}{3}r^{3}\frac{d\bar\jmath^{2}}{dr}\varpi^{2},
\label{eq:m0}\\
\frac{dp_{0}^{*}}{dr}=&-m_{0}\frac{1+8\pi r^{2}P}{(r-2m)^{2}}
-\frac{4\pi(\varepsilon+P)r^{2}}{r-2m}\,p_{0}^{*}
+\frac{\bar\jmath^{2}r^{4}\varpi'^{2}}{12(r-2m)}\nonumber\\
&+\frac{1}{3}\frac{d}{dr}\left[\frac{r^{3}\bar\jmath^{2}\varpi^{2}}{r-2m}\right].
\label{eq:p0dot}
\end{align}
The first is the $tt$ equation: the change of the enclosed mass has a
contribution from the change of the density and two from the rotational
energy of fluid and field.  The terms in $\varpi$ act as sources, so $m_{0}$
and $p_{0}^{*}$ are of order $\Omega^{2}$.  For a sequence at fixed central
density, $m_{0}=p_{0}^{*}=0$ at $r=0$.

Outside the star $\bar\jmath=1$ and $\varpi'=6J/r^{4}$, and \Eqref{eq:m0}
becomes $m_{0}'=3J^{2}/r^{4}$, with solution
\begin{align}
m_{0}&=\delta M-\frac{J^{2}}{r^{3}},\nonumber\\
h_{0}&=-\frac{\delta M}{r-2M}+\frac{J^{2}}{r^{3}(r-2M)}\qquad(r>R),
\label{eq:m0ext}
\end{align}
where the form of $h_{0}$ follows from the $rr$ equation.  At large $r$,
$-g_{tt}=(1-2M/r)(1+2h_{0})=1-2(M+\delta M)/r+\mathcal{O}(r^{-3})$ and
$g_{rr}=1+2(M+\delta M)/r+\dots$, so $\delta M$ is the change of the
gravitational mass.  Continuity of $m_{0}$ at $R$ gives
\begin{equation}
\delta M=m_{0}(R)+\frac{J^{2}}{R^{3}},
\end{equation}
and the total mass is $M=M_{\rm TOV}+\delta M$.  In the exterior metric of
\Eqref{eq:exterior} the constant $\delta M$ has been absorbed into $M$, so
that $m_{0}=-J^{2}/r^{3}$ and $h_{0}=J^{2}/[r^{3}(r-2M)]$ there.  The
constant $h_{0c}$ of \Eqref{eq:p0} follows from the continuity of $h_{0}$,
but it is not needed for any quantity used in this paper.

\subsection{Quadrupole deformation}

The $P_{2}$ parts of the field equations involve $h_{2}$, $m_{2}$ and $k_{2}$.
The $\theta\theta-\phi\phi$ component is algebraic and expresses $m_{2}$ through
$h_{2}$ and the rotation; outside the star it gives
$m_{2}=(r-2M)(-h_{2}+6J^{2}/r^{4})$.  With $v_{2}\equiv h_{2}+k_{2}$, the
remaining components give two first-order equations \citep{Hartle1967},
\begin{align}
\frac{dv_{2}}{dr}=&-\nu'h_{2}+\left(\frac{1}{r}+\frac{\nu'}{2}\right)
\left[\frac{1}{6}r^{4}\bar\jmath^{2}\varpi'^{2}-\frac{1}{3}r^{3}\frac{d\bar\jmath^{2}}{dr}\varpi^{2}\right],
\label{eq:v2}\\
\frac{dh_{2}}{dr}=&-\nu'h_{2}+\frac{r\,h_{2}}{(r-2m)\nu'}\left[8\pi(\varepsilon+P)-\frac{4m}{r^{3}}\right]\nonumber\\
&-\frac{4v_{2}}{r(r-2m)\nu'}\nonumber\\
&+\frac{1}{6}\left[\frac{r\nu'}{2}-\frac{1}{(r-2m)\nu'}\right]r^{3}\bar\jmath^{2}\varpi'^{2}\nonumber\\
&-\frac{1}{3}\left[\frac{r\nu'}{2}+\frac{1}{(r-2m)\nu'}\right]r^{2}\frac{d\bar\jmath^{2}}{dr}\varpi^{2}.
\label{eq:h2}
\end{align}

At the centre, regularity requires $h_{2}\simeq Ar^{2}$ and $v_{2}\simeq Br^{4}$.  The
rotation terms are of higher order in $r$, and \Eqref{eq:v2} with
\Eqref{eq:nucentre} gives
\begin{equation}
B=-\frac{2\pi}{3}\left(\varepsilon_{c}+3P_{c}\right)A .
\end{equation}
The general regular solution is therefore $h_{2}=h_{2}^{\rm P}+Ah_{2}^{\rm H}$,
$v_{2}=v_{2}^{\rm P}+Av_{2}^{\rm H}$, where the particular solution
$(h_{2}^{\rm P},v_{2}^{\rm P})$ solves equations (\ref{eq:v2})--(\ref{eq:h2})
with zero initial values, and the homogeneous solution
$(h_{2}^{\rm H},v_{2}^{\rm H})$ solves them without the $\varpi$ terms,
starting from $h_{2}^{\rm H}=r^{2}$, $v_{2}^{\rm H}=-\tfrac{2\pi}{3}(\varepsilon_{c}+3P_{c})r^{4}$.

The exterior solution (Appendix~\ref{app:exterior}) has one
free constant $K$.  Continuity of $h_{2}$ and $v_{2}$ at $r=R$,
\begin{align}
h_{2}^{\rm P}(R)+A\,h_{2}^{\rm H}(R)&=\frac{J^{2}}{MR^{3}}\left(1+\frac{M}{R}\right)+K\,Q_{2}^{2}(\xi_{R}),\nonumber\\
v_{2}^{\rm P}(R)+A\,v_{2}^{\rm H}(R)&=-\frac{J^{2}}{R^{4}}+K\,\frac{2M}{\sqrt{R(R-2M)}}\,Q_{2}^{1}(\xi_{R}),
\label{eq:h2match}
\end{align}
with $\xi_{R}=R/M-1$, is a linear system for $A$ and $K$.  The quadrupole
moment is then given by \Eqref{eq:Q} (Appendix~\ref{app:moments}).  Because
all source terms are quadratic in $\varpi\propto\Omega$, $h_{2}$, $v_{2}$, $K$ and $Q$
are proportional to $\Omega^{2}$ at fixed central density.

\subsection{Surface of the rotating star}

A fluid element on the surface of the nonrotating star is displaced to
$r=R+\xi(\theta)$, where the total pressure vanishes:
$P(R+\xi)+(\varepsilon+P)p^{*}=0$.  To first order in $\xi$,
\begin{equation}
\xi=-\frac{(\varepsilon+P)p^{*}}{P'}=p^{*}\frac{r(r-2m)}{m+4\pi r^{3}P},
\end{equation}
which at the surface, where $P=0$ and $m=M$, is
$\xi=p^{*}(R)\,R(R-2M)/M$.
With $p^{*}=p_{0}^{*}+p_{2}^{*}P_{2}$, the surface is
$r(\theta)=R+\xi_{0}+\xi_{2}P_{2}(\cos\theta)$, with $\xi_{0}$ from $p_{0}^{*}(R)$ and
$\xi_{2}$ from \Eqref{eq:p2} at $r=R$.  At the equator and the pole
($P_{2}=-\tfrac{1}{2}$ and 1),
\begin{equation}
R_{\rm eq}=R+\xi_{0}-\tfrac{1}{2}\xi_{2},\qquad R_{\rm pol}=R+\xi_{0}+\xi_{2}.
\end{equation}
The circumference of the equator is $2\pi\sqrt{g_{\phi\phi}}$ at $\theta=\pi/2$, and
from \Eqref{eq:htmetric}
\begin{equation}
R_{\rm eq,circ}=R_{\rm eq}\left[1+k_{2}(R)P_{2}(0)\right]=R_{\rm eq}\left[1-\tfrac{1}{2}k_{2}(R)\right].
\end{equation}
The coordinate radius $R_{\rm eq}$ is the inner edge of a disc that touches
the star, in the same coordinate as the exterior metric used for the
orbits.

\subsection{Stars of given mass and spin}

For a target total mass $M$ and spin frequency $\nu_s$ we solve
$M_{\rm TOV}(\varepsilon_c)+\delta M(\varepsilon_c,\nu_s)=M$ on the stable branch
by a secant iteration in $\ln\varepsilon_c$.  For the fits we compute one
model at each of 64 central densities, from $0.9\,\mathrm{M}_\odot$ to the
maximum mass, at a reference spin frequency $\nu_{\rm ref}$, and use the
scalings of this appendix at fixed $\varepsilon_c$,
\begin{equation}
J\propto\nu_s,\qquad \delta M,\;Q,\;\xi_{0},\;\xi_{2}\propto\nu_s^{2},
\end{equation}
to obtain all quantities at any $\nu_s$, after which $M(\varepsilon_c,\nu_s)$ is
inverted by interpolation.  These scalings are exact in the HT approximation.

\section{Exterior spacetime and multipole moments}
\label{app:exterior}

\subsection{Vacuum solutions}

Outside the star $\varepsilon=P=0$, $m=M$ and $\nu'=2M/[r(r-2M)]$.  The
frame dragging and the monopole deformation were solved above.  For the
quadrupole sector the sources in equations (\ref{eq:v2})--(\ref{eq:h2})
reduce to the $\varpi'^{2}=36J^{2}/r^{8}$ terms.  A particular solution is
\begin{equation}
h_{2}^{\rm P}=\frac{J^{2}}{Mr^{3}}\left(1+\frac{M}{r}\right),\qquad v_{2}^{\rm P}=-\frac{J^{2}}{r^{4}},
\end{equation}
as direct substitution shows.  For the homogeneous part we eliminate
$v_{2}$: \Eqref{eq:h2} gives $v_{2}$ in terms of $h_{2}$ and $h_{2}'$, and
inserting it into \Eqref{eq:v2} gives a second-order equation for $h_{2}$.
In the variable $\xi=r/M-1$ it is
\begin{equation}
(1-\xi^{2})\frac{d^{2}h_{2}}{d\xi^{2}}-2\xi\frac{dh_{2}}{d\xi}+\left[6-\frac{4}{1-\xi^{2}}\right]h_{2}=0,
\label{eq:legendre}
\end{equation}
the associated Legendre equation of degree 2 and order 2.  Its solutions are
$P_{2}^{2}(\xi)\propto\xi^{2}-1$, which grows at large $r$ and is excluded by
asymptotic flatness, and the function of the second kind $Q_{2}^{2}(\xi)$.
Integrating $v_{2}'=-\nu'h_{2}$ then gives the accompanying $v_{2}$.  The
result is \Eqref{eq:exterior},
\begin{align}
h_{2}&=\frac{J^{2}}{Mr^{3}}\left(1+\frac{M}{r}\right)+K\,Q_{2}^{2}(\xi),\nonumber\\
v_{2}&=-\frac{J^{2}}{r^{4}}+K\frac{2M}{\sqrt{r(r-2M)}}\,Q_{2}^{1}(\xi),
\end{align}
with
\begin{align}
Q_{2}^{1}(\xi)&=\sqrt{\xi^{2}-1}\left[\frac{3\xi^{2}-2}{\xi^{2}-1}-\frac{3}{2}\xi\ln\frac{\xi+1}{\xi-1}\right],\nonumber\\
Q_{2}^{2}(\xi)&=\frac{3}{2}(\xi^{2}-1)\ln\frac{\xi+1}{\xi-1}-\frac{3\xi^{3}-5\xi}{\xi^{2}-1},
\end{align}
for $\xi>1$, in the sign convention of \citet{HartleThorne1968}, in which both
are positive.  For large $r$, with $s=M/r$,
\begin{equation}
Q_{2}^{2}=\tfrac{8}{5}s^{3}+\tfrac{24}{5}s^{4}+\mathcal{O}(s^{5}),\qquad
Q_{2}^{1}=\tfrac{2}{5}s^{3}+\tfrac{6}{5}s^{4}+\mathcal{O}(s^{5}).
\label{eq:Qasym}
\end{equation}
We have checked by symbolic computation that these functions satisfy
equations (\ref{eq:v2})--(\ref{eq:h2}) in vacuum, and that the change of
variable turns them into \Eqref{eq:legendre}.  Because the closed forms lose
precision through cancellation at large $\xi$, we evaluate them by the
convergent series in $\xi^{-2}$ for $\xi\ge3$.

\subsection{Multipole moments}
\label{app:moments}

By \Eqref{eq:Qasym}, at large $r$
\begin{equation}
h_{2}\simeq\left(\frac{J^{2}}{M}+\frac{8}{5}KM^{3}\right)\frac{1}{r^{3}}\equiv\frac{Q_{\rm HT}}{r^{3}},
\end{equation}
and
\begin{align}
-g_{tt}&=\left(1-\frac{2M}{r}\right)\left(1+2h_{2}P_{2}\right)+\dots\nonumber\\
&=1-\frac{2M}{r}+\frac{2Q_{\rm HT}}{r^{3}}P_{2}+\mathcal{O}(r^{-4}).
\end{align}
For a Newtonian body with quadrupole moment $Q$ the potential is
$\Phi=-M/r-QP_{2}/r^{3}$, and $-g_{tt}\simeq1+2\Phi$.  An oblate body has more
mass near the equator, where $P_{2}=-\tfrac{1}{2}$, so its potential is
deeper there and $Q<0$; this is the Geroch--Hansen sign convention, in
which the Kerr metric has $Q=-J^{2}/M$.  Comparison gives
$Q=-Q_{\rm HT}$, which is \Eqref{eq:Q}.  The spin $J$ was identified in the
same way from $g_{t\phi}$.  The dimensionless moments are $j=J/M^{2}$ and
$q=Q/M^{3}$.

\subsection{Kerr limit}

For a Kerr black hole of mass $M$ and specific angular momentum $a=J/M$, the
quadrupole is $Q=-Ma^{2}=-J^{2}/M$, so $Q_{\rm HT}=J^{2}/M$ and $K=0$.  The
HT exterior with $K=0$ must then be the Kerr metric expanded to second order
in $a$.  Writing the Boyer--Lindquist coordinates $(r_{\rm BL},\theta_{\rm BL})$ as
\begin{align}
r_{\rm BL}&=r-\frac{a^{2}}{2r}\Bigg[\left(1+\frac{2M}{r}\right)\left(1-\frac{M}{r}\right)\nonumber\\
&\qquad-\cos^{2}\theta\left(1-\frac{2M}{r}\right)\left(1+\frac{3M}{r}\right)\Bigg],\nonumber\\
\theta_{\rm BL}&=\theta-\frac{a^{2}}{2r^{2}}\left(1+\frac{2M}{r}\right)\sin\theta\cos\theta,
\end{align}
\citep{HartleThorne1968}, we have verified symbolically that every component of
the Kerr metric, including the induced $g_{r\theta}$, agrees with
\Eqref{eq:gcomp} to $\mathcal{O}(a^{2})$.  This is the analytic basis of the
statement that the HT frequencies with $q=-j^{2}$ reproduce the Kerr
frequencies to $\mathcal{O}(j^{2})$.

\subsection{Metric components used for the orbits}

Collecting the exterior functions, and absorbing $\delta M$ into $M$,
\begin{align}
g_{tt}&=-\left(1-\frac{2M}{r}\right)\left[1+2h_{2}P_{2}+\frac{2J^{2}}{r^{4}}\frac{2\cos^{2}\theta-1}{1-2M/r}\right],\nonumber\\
g_{t\phi}&=-\frac{2J}{r}\sin^{2}\theta,\qquad g_{\theta\theta}=r^{2}\left(1+2k_{2}P_{2}\right),\nonumber\\
g_{\phi\phi}&=r^{2}\sin^{2}\theta\left(1+2k_{2}P_{2}\right),\nonumber\\
g_{rr}&=\left(1-\frac{2M}{r}\right)^{-1}\nonumber\\
&\quad\times\left[1-2\left(h_{2}-\frac{6J^{2}}{r^{4}}\right)P_{2}-\frac{2J^{2}}{r^{4}(1-2M/r)}\right],
\label{eq:gcomp}
\end{align}
with $k_{2}=v_{2}-h_{2}$.  The $J^{2}$ terms of $g_{tt}$ collect $h_{0}$ and the
$\omega^{2}r^{2}\sin^{2}\theta$ term of $(d\phi-\omega\,dt)^{2}$; the $J^{2}$ terms of
$g_{rr}$ collect $m_{0}$ and $m_{2}$.  The metric depends on $(M,J,Q)$ only.

\section{Circular orbits and epicyclic frequencies}
\label{app:epicyclic}

\subsection{Conserved quantities and the effective potential}

Consider a timelike geodesic in a stationary, axisymmetric metric that is
symmetric under reflection in the equatorial plane, with nonzero components
$g_{tt}$, $g_{t\phi}$, $g_{\phi\phi}$, $g_{rr}$ and $g_{\theta\theta}$ depending on $(r,\theta)$.  With
$\mathcal{L}=\tfrac{1}{2}g_{\mu\nu}\dot x^{\mu}\dot x^{\nu}$ and a dot denoting $d/d\tau$, the
coordinates $t$ and $\phi$ are cyclic, and the specific energy and angular
momentum
\begin{equation}
E=-u_{t}=-(g_{tt}\dot t+g_{t\phi}\dot\phi),\qquad L=u_{\phi}=g_{t\phi}\dot t+g_{\phi\phi}\dot\phi
\end{equation}
are conserved.  Inverting the $(t,\phi)$ block, with
$D=g_{tt}g_{\phi\phi}-g_{t\phi}^{2}<0$,
\begin{equation}
\dot t=\frac{g_{\phi\phi}E+g_{t\phi}L}{-D},\qquad
\dot\phi=-\frac{g_{t\phi}E+g_{tt}L}{-D},
\end{equation}
and the $(t,\phi)$ part of $g^{\mu\nu}u_{\mu}u_{\nu}$ is
\begin{equation}
\mathcal{H}(r,\theta;E,L)=\frac{g_{\phi\phi}E^{2}+2g_{t\phi}EL+g_{tt}L^{2}}{D}.
\end{equation}
The normalisation $u\cdot u=-1$ becomes
\begin{equation}
g_{rr}\dot r^{2}+g_{\theta\theta}\dot\theta^{2}=-1-\mathcal{H}(r,\theta;E,L)\equiv V(r,\theta),
\label{eq:Veff}
\end{equation}
so motion is allowed where $V\ge0$.  Eliminating the cyclic coordinates by a
Routh transformation gives the reduced Lagrangian
$\mathcal{R}=\tfrac{1}{2}(g_{rr}\dot r^{2}+g_{\theta\theta}\dot\theta^{2})-\tfrac{1}{2}\mathcal{H}$, whose
Euler--Lagrange equations are
\begin{align}
\frac{d}{d\tau}\left(g_{rr}\dot r\right)&=\tfrac{1}{2}\partial_{r}g_{rr}\,\dot r^{2}+\tfrac{1}{2}\partial_{r}g_{\theta\theta}\,\dot\theta^{2}-\tfrac{1}{2}\partial_{r}\mathcal{H},\nonumber\\
\frac{d}{d\tau}\left(g_{\theta\theta}\dot\theta\right)&=\tfrac{1}{2}\partial_{\theta}g_{rr}\,\dot r^{2}+\tfrac{1}{2}\partial_{\theta}g_{\theta\theta}\,\dot\theta^{2}-\tfrac{1}{2}\partial_{\theta}\mathcal{H}.
\label{eq:routh}
\end{align}

\subsection{Circular equatorial orbits}

On the equator $\partial_{\theta}\mathcal{H}=0$ by reflection symmetry.  A circular orbit
of radius $r$ has $\dot r=\dot\theta=0$ for all $\tau$, which by
\Eqref{eq:routh} requires $\partial_{r}\mathcal{H}=0$, and by \Eqref{eq:Veff} $V=0$.
Equivalently, the radial component of the geodesic equation,
$\Gamma^{r}{}_{\mu\nu}u^{\mu}u^{\nu}=-\tfrac{1}{2}g^{rr}\partial_{r}g_{\mu\nu}u^{\mu}u^{\nu}=0$, with
$u^{\phi}=\Omega_{\phi}u^{t}$, gives
\begin{equation}
\partial_{r}g_{tt}+2\Omega_{\phi}\,\partial_{r}g_{t\phi}+\Omega_{\phi}^{2}\,\partial_{r}g_{\phi\phi}=0 .
\end{equation}
Its prograde root is \Eqref{eq:omphi}.  The normalisation then gives
\begin{equation}
u^{t}=\left[-\left(g_{tt}+2\Omega_{\phi}g_{t\phi}+\Omega_{\phi}^{2}g_{\phi\phi}\right)\right]^{-1/2},
\end{equation}
and $E=-(g_{tt}+g_{t\phi}\Omega_{\phi})u^{t}$, $L=(g_{t\phi}+g_{\phi\phi}\Omega_{\phi})u^{t}$.  For the
Schwarzschild metric these give $\Omega_{\phi}^{2}=M/r^{3}$, and for Kerr
$\Omega_{\phi}=M^{1/2}/(r^{3/2}+aM^{1/2})$.

\subsection{Small oscillations}

Let $r=r_{0}+\delta r$ and $\theta=\pi/2+\delta\theta$ with $E$ and $L$ those of the
circular orbit at $r_{0}$.  The terms in $\dot r^{2}$ and $\dot\theta^{2}$ in
\Eqref{eq:routh} are of second order, $\partial_{r}\mathcal{H}$ and $\partial_{\theta}\mathcal{H}$ vanish on
the circular orbit, and $\partial_{r}\partial_{\theta}\mathcal{H}=0$ on the equator by reflection
symmetry.  To first order,
\begin{equation}
g_{rr}\,\delta\ddot r=-\tfrac{1}{2}\mathcal{H}_{,rr}\,\delta r,\qquad
g_{\theta\theta}\,\delta\ddot\theta=-\tfrac{1}{2}\mathcal{H}_{,\theta\theta}\,\delta\theta ,
\end{equation}
two uncoupled harmonic oscillators with proper-time angular frequencies
\begin{equation}
\kappa_{r}^{2}=\frac{\mathcal{H}_{,rr}}{2g_{rr}},\qquad
\kappa_{\theta}^{2}=\frac{\mathcal{H}_{,\theta\theta}}{2g_{\theta\theta}},
\end{equation}
evaluated at $(r_{0},\pi/2)$.  A distant observer measures time $t$, and
$dt/d\tau=u^{t}$, which gives \Eqref{eq:nus}.  For the Schwarzschild metric
$\kappa_{r}^{2}/(u^{t})^{2}=\Omega_{\phi}^{2}(1-6M/r)$ and $\kappa_{\theta}/u^{t}=\Omega_{\phi}$, so the
vertical frequency equals the orbital frequency and orbits do not precess
nodally.  In our implementation the second derivatives are taken by
fourth-order finite differences, which reproduce these Schwarzschild
results to $\sim10^{-10}$.

\subsection{Innermost stable circular orbit}

A circular orbit is stable against radial perturbations if $\kappa_{r}^{2}>0$.
As $r$ decreases, $\kappa_{r}^{2}$ changes sign at the ISCO, where also
$dE/dr=dL/dr=0$.  We take the ISCO as the outermost zero of $\kappa_{r}^{2}$
and the inner edge of the disc as $r_{\rm in}=\max(R_{\rm eq},r_{\rm ISCO})$.

\section{Weak-field expansions}
\label{app:weak}

To interpret the results we expand the frequencies for large $x=r/M$,
keeping for each multipole only its leading term.  At this order the
contributions of $M$, $J$ and $Q$ add linearly.

For the mass and the spin we use the Kerr metric, for which in
Boyer--Lindquist coordinates \citep{Okazaki1987}
\begin{align}
\frac{\Omega_{r}^{2}}{\Omega_{\phi}^{2}}&=1-\frac{6}{x}+\frac{8j}{x^{3/2}}-\frac{3j^{2}}{x^{2}},\qquad
\frac{\Omega_{\theta}^{2}}{\Omega_{\phi}^{2}}=1-\frac{4j}{x^{3/2}}+\frac{3j^{2}}{x^{2}},
\end{align}
where $\Omega_{r,\theta}=\kappa_{r,\theta}/u^{t}$.  The terms linear in $j$ come from
the gravitomagnetic potential $g_{t\phi}=-2J\sin^{2}\theta/r$ and hold for any
spinning body; the $j^{2}$ terms are those of the Kerr quadrupole
$q=-j^{2}$.

For the quadrupole, a Newtonian body with potential
$\Phi=-M/r-QP_{2}(\cos\theta)/r^{3}$ has, on the equator,
\begin{align}
\Omega_{\phi}^{2}&=\frac{1}{r}\frac{\partial\Phi}{\partial r}=\frac{M}{r^{3}}-\frac{3}{2}\frac{Q}{r^{5}},\nonumber\\
\kappa_{r}^{2}&=\frac{1}{r^{3}}\frac{d}{dr}\left(r^{4}\Omega_{\phi}^{2}\right)=\frac{M}{r^{3}}+\frac{3}{2}\frac{Q}{r^{5}},\nonumber\\
\kappa_{\theta}^{2}&=\frac{1}{r}\frac{\partial\Phi}{\partial r}+\frac{1}{r^{2}}\frac{\partial^{2}\Phi}{\partial\theta^{2}}=\Omega_{\phi}^{2}-\frac{3Q}{r^{5}},
\end{align}
using $\partial_{\theta}^{2}P_{2}=3$ at $\theta=\pi/2$.  To first order in $q=Q/M^{3}$,
\begin{equation}
\frac{\kappa_{r}^{2}}{\Omega_{\phi}^{2}}=1+\frac{3q}{x^{2}},\qquad
\frac{\kappa_{\theta}^{2}}{\Omega_{\phi}^{2}}=1-\frac{3q}{x^{2}} .
\end{equation}
For $q=-j^{2}$ these are exactly the $j^{2}$ terms of the Kerr expressions,
so the two cases combine into \Eqref{eq:weak}, valid for independent $j$
and $q$ at this order.

Taking square roots gives the precession frequencies,
\begin{align}
\nu_{\rm per}&\simeq\nu_{\phi}\left(\frac{3}{x}+\frac{9}{2x^{2}}-\frac{4j}{x^{3/2}}-\frac{3}{2}\frac{q}{x^{2}}+\dots\right),\\
\nu_{\rm nod}&\simeq\nu_{\phi}\left(\frac{2j}{x^{3/2}}+\frac{3}{2}\frac{q}{x^{2}}+\dots\right).
\end{align}
With $\nu_{\phi}\simeq(2\pi M)^{-1}x^{-3/2}$ at leading order, the spin term of
$\nu_{\rm nod}$ is $j/(\pi Mx^{3})=J/(\pi r^{3})=\omega/2\pi$, the Lense--Thirring
frequency of \Eqref{eq:nod}.  The quadrupole term is the classical
precession of an orbit around an oblate body, retrograde for $q<0$.  Their
ratio, with $q=-\bar Qj^{2}$, is \Eqref{eq:ratio}.  The expansions are used
only for interpretation; all fits use the exact frequencies of
Appendix~\ref{app:epicyclic}.  Near the ISCO they are accurate to about
10~per~cent.

\section{Geodesics, hot spot and light curve}
\label{app:orbit}

\subsection{Geodesic equations and initial data}

We integrate $\dot x^{\mu}=u^{\mu}$ and $\dot u^{\mu}=-\Gamma^{\mu}{}_{\alpha\beta}u^{\alpha}u^{\beta}$ in the
metric of \Eqref{eq:gcomp}, with Christoffel symbols computed from the
analytic metric by fourth-order finite differences.  The orbit starts at
$(r_{0},\pi/2,0)$ with coordinate velocities
\begin{align}
\frac{d\phi}{dt}&=\Omega_{\phi}(r_{0})\cos\iota,\qquad
\frac{d\theta}{dt}=\Omega_{\phi}(r_{0})\sin\iota,\nonumber\\
\frac{dr}{dt}&=e\,\Omega_{\phi}(r_{0})\,r_{0},
\end{align}
where $\iota$ is the tilt of the orbital plane and $e$ the radial kick, and
$u^{t}$ follows from $u\cdot u=-1$.  For $e=\iota=0$ this is the circular orbit
of Appendix~\ref{app:epicyclic}; for small $e$ and $\iota$ the orbit oscillates
about $r_{0}$ and $\theta=\pi/2$ with the epicyclic frequencies at $r_{0}$.  The
conserved $E$ and $L_{z}$ and the norm $u\cdot u$ are monitored along the
orbit (\secref{sec:orbits}).

\subsection{Measuring the precession}

From $(r,\theta,\phi)$ we form flat-space Cartesian coordinates $\bm{x}$ and
velocities $\bm{v}=d\bm{x}/dt$, and the orbital angular momentum direction
$\bm{l}\propto\bm{x}\times\bm{v}$.  The ascending node lies along
$\hat{\bm z}\times\bm{l}$, so its longitude is $\Omega_{\rm N}=\arctan(l_{x},-l_{y})$, and
$\nu_{\rm nod}=\dot\Omega_{\rm N}/2\pi$.  The pericentres are the local minima of $r(t)$;
the longitude of the pericentre advances at the rate
$2\pi(\nu_{\phi}-\nu_{r})=2\pi\nu_{\rm per}$.  Both rates are obtained by linear fits
to the angles as functions of time.

\subsection{Redshift and flux of a hot spot}

We treat the emitter as a point source on the orbit that emits
isotropically in its rest frame, and the photons as moving on straight
lines in the direction $\hat{\bm n}=(\sin i,0,\cos i)$ of a distant observer at
inclination $i$.  This neglects light bending and the dragging of photons,
which change the amplitudes of the modulation but not its frequencies.  A
photon received at infinity with unit energy has $k_{t}=-1$ and spatial
components $k_{j}=n_{j}$ in this approximation.  The energy measured by the
emitter is $-k_{\mu}u^{\mu}=u^{t}-n_{j}u^{j}=u^{t}(1-\hat{\bm n}\cdot\bm v)$, so the ratio
of observed to emitted photon energy is
\begin{equation}
g=\frac{1}{u^{t}\left(1-\hat{\bm n}\cdot\bm v\right)} .
\end{equation}
The factor $1/u^{t}$ contains the gravitational and transverse Doppler
shifts, and $(1-\hat{\bm n}\cdot\bm v)^{-1}$ the longitudinal Doppler shift.
Because $I_{\nu}/\nu^{3}$ is invariant along rays, the bolometric intensity
scales as $g^{4}$, and so does the bolometric flux of a compact source,
$F\propto g^{4}$, as for a relativistically moving source in flat space.  A photon emitted at
coordinate time $t$ from position $\bm x$ arrives at
\begin{equation}
t_{\rm obs}=t-\hat{\bm n}\cdot\bm x
\end{equation}
up to a constant (Roemer delay).  We resample $F(t_{\rm obs})$ on a uniform
grid and compute its power spectrum with a Hann window.

\subsection{Which frequencies appear}

The origin of the peaks in \figref{fig:psd} follows from an expansion in
the small quantities $e$ and $\iota$.  For the tilted orbit the vertical
displacement is $z\simeq r\iota\sin(2\pi\nu_{\theta}t)$, because the angle between the
particle and the precessing node advances at $2\pi(\nu_{\phi}-\nu_{\rm nod})=2\pi\nu_{\theta}$.
With $\beta=\hat{\bm n}\cdot\bm v=v_{x}\sin i+v_{z}\cos i$,
\begin{align}
v_{x}&\simeq-r\Omega_{\phi}\sin(2\pi\nu_{\phi}t),\qquad
v_{z}\simeq r\iota\,\Omega_{\theta}\cos(2\pi\nu_{\theta}t),\nonumber\\
F&\propto(u^{t})^{-4}\left(1+4\beta+10\beta^{2}+20\beta^{3}+35\beta^{4}+\dots\right).
\end{align}
The term linear in $\beta$ gives peaks at $\nu_{\phi}$, with amplitude
$\propto r\Omega_{\phi}\sin i$, and at $\nu_{\theta}$, with amplitude $\propto\iota\,r\Omega_{\theta}\cos i$.
The term $10\beta^{2}$ contains $2v_{x}v_{z}\sin i\cos i$, whose product of
oscillations at $\nu_{\phi}$ and $\nu_{\theta}$ has a component at
$\nu_{\phi}-\nu_{\theta}=\nu_{\rm nod}$, with amplitude $\propto\iota\,(r\Omega_{\phi})^{2}\sin2i$.  The radial
epicycle, $r=r_{0}[1+\mathcal{O}(e)\cos(2\pi\nu_{r}t)]$, modulates $u^{t}$ and $|\bm v|$
at $\nu_{r}$ and, multiplied by the orbital term, gives sidebands at
$\nu_{\phi}\pm\nu_{r}$; the lower sideband is $\nu_{\rm per}$.  A component at
$2\nu_{\rm nod}$ needs $v_{x}^{2}v_{z}^{2}$, which appears first in the $\beta^{4}$ term with
amplitude $\propto\iota^{2}(r\Omega_{\phi})^{4}\sin^{2}2i$.  It is smaller than the
$\nu_{\rm nod}$ term by a further factor of order $\iota(r\Omega_{\phi})^{2}$, which is why a
single orbiting spot produces a weak peak at $2\nu_{\rm nod}$.  The Roemer delay
modulates the phase at the same frequencies and does not add new ones.
Since $r\Omega_{\phi}\simeq0.4$ for the orbit of \secref{sec:orbits}, the series is
not rapidly convergent, and this argument predicts which frequencies appear
and how their amplitudes scale, not their exact values.

\section{Likelihood and radius marginalisation}
\label{app:like}

\subsection{Likelihood}

A source has $n$ sets of simultaneous QPOs.  Set $i$ contains measured
frequencies $d_{ik}$ with errors $\sigma_{ik}$ in the channels
$k\in\{U,L,{\rm LF}\}$ that were observed.  For given model parameters
$\bm\theta$ (for example $M$ and $\nu_s$ with an EOS, or $M$, $j$ and $\bar Q$ without
one), the RPM predicts frequencies $m_{k}(x;\bm\theta)$ along the radius
$x=r/M$ (\Eqref{eq:rpm}).  The radius of each set is unknown and is
marginalised with a uniform prior on $[x_{\rm in},x_{\rm max}]$, with
$x_{\rm in}=r_{\rm in}/M$ and $x_{\rm max}=25$:
\begin{align}
\mathcal{L}(\bm\theta,\bm\eta)=&\prod_{i=1}^{n}\frac{1}{x_{\rm max}-x_{\rm in}}
\int_{x_{\rm in}}^{x_{\rm max}}dx\nonumber\\
&\times\prod_{k}\frac{1}{\sqrt{2\pi s_{ik}^{2}}}
\exp\left[-\frac{(d_{ik}-m_{k}(x;\bm\theta))^{2}}{2s_{ik}^{2}}\right].
\label{eq:like}
\end{align}
The variance $s_{ik}^{2}=\sigma_{ik}^{2}+(\eta_{k}d_{ik})^{2}$ includes a fractional
intrinsic scatter, which absorbs the QPO widths and any inadequacy of the
model.  In the two-scatter model $\eta_{U}=\eta_{L}=\eta_{\rm kHz}$ and
$\eta_{\rm LF}$ is separate; in the one-scatter model all channels share one
$\eta$.

\subsection{Exact marginalisation over the radius}

The model curves are tabulated on a grid $x_{0}<x_{1}<\dots<x_{K}$ and
interpolated linearly.  On the interval $[x_{a},x_{a+1}]$, with
$t=x-x_{a}\in[0,h_{a}]$, each curve is $m_{k}=m_{k,a}+b_{k}t$.  With residuals
$\rho_{k}=d_{ik}-m_{k,a}$ and weights $w_{k}=s_{ik}^{-2}$ the exponent of
\Eqref{eq:like} is $-\tfrac{1}{2}(At^{2}-2Bt+C)$, with
\begin{equation}
A=\sum_{k}w_{k}b_{k}^{2},\qquad B=\sum_{k}w_{k}b_{k}\rho_{k},\qquad C=\sum_{k}w_{k}\rho_{k}^{2}.
\end{equation}
Completing the square,
\begin{align}
\int_{0}^{h_{a}}e^{-\frac{1}{2}(At^{2}-2Bt+C)}dt
=&\sqrt{\frac{2\pi}{A}}\;e^{-\frac{1}{2}\left(C-B^{2}/A\right)}\nonumber\\
&\times\left[\Phi(u_{a}+\sqrt{A}h_{a})-\Phi(u_{a})\right],
\end{align}
with $u_{a}=-B/\sqrt{A}$ and $\Phi$ the standard normal distribution function.
The integral over $x$ is the sum over intervals.  This is exact for the
interpolated curves and remains accurate when the measurement errors are
smaller than the change of the model between grid points, as for the
lower kHz QPO of 4U 1728$-$34 ($\sigma\simeq1$~Hz); averaging the integrand over
grid points would then be undersampled.  We have checked the closed form
against direct quadrature to $10^{-8}$ in $\ln\mathcal{L}$.

\subsection{Priors, evidence and derived quantities}

With an EOS, the mass has a uniform prior on 1.0--$2.6\,\mathrm{M}_\odot$; the
spin frequency is fixed at the burst-oscillation value for the atoll
sources and uniform on 200--750~Hz for the Z sources.  Without an EOS, $j$
and $\bar Q$ have uniform priors on 0--0.45 and 1--12, and the mass prior is
uniform up to 2.6~$\mathrm{M}_\odot$ for the atoll sources and 3.2~$\mathrm{M}_\odot$ for
the Z sources.  The scatters have
log-uniform priors on $10^{-4}$--0.2.  The posterior is
$p(\bm\theta,\bm\eta\,|\,\bm d)=\mathcal{L}\,\pi/Z$ with the evidence
$Z=\int\mathcal{L}\,\pi\,d\bm\theta\,d\bm\eta$.  For the fits with an EOS we evaluate $Z$ and
the posterior by quadrature on a grid in $(M,\nu_s,\bm\eta)$; for the fits without
an EOS we use nested sampling \citep{Ashton2019,Speagle2020}.  We quote
medians and 68~per~cent intervals of the marginal posteriors.  The derived
quantities follow from the star: with an EOS, $(M,\nu_s)\mapsto(j,q,I)$ by
Appendix~\ref{app:interior}; without one, $I=J/\Omega=jM^{2}/(2\pi\nu_s)$ in units
with $G=c=1$, that is $I=jGM^{2}/(2\pi c\,\nu_s)$.

\bsp
\label{lastpage}
\end{document}